\documentclass[reprint,aps,prf,floatfix,longbibliography,10pt]{revtex4-1}
\newcommand{\RomanNumeralCaps}[1]{\MakeUppercase{\romannumeral #1}}

\usepackage{amsmath,amssymb,color,latexsym,natbib,graphicx}
\usepackage{dcolumn, enumerate}
\usepackage{bm}
\usepackage{soul}

\DeclareMathAlphabet\mathbfcal{OMS}{cmsy}{b}{n}
 
\newcommand{\p} {\partial}
\def\eg{{\it e.g.}\ } 

\def\uu{{\bf u}}

\def\d_M{{\bf d_M}}

\def\bb{{\bf b}}

\def\xx{{\bf x}}

\def\EE{{\bf E}}

\def\Bz{{\bf B_0}}
\def\Jc{\frac{\JJ}{n e}}

\def\bomega{{\boldsymbol \omega}}
\def \pmbmath{\mathpalette\pmbmathaux}
\def \pmbmathaux#1#2{
	\pmbtext{$#1#2$}}
\def \pmbtext#1{\leavevmode
	\setbox0\hbox{#1}
	\kern0,4pt \copy0 \kern-\wd0
	\kern-0,2pt \raise0,3pt \box0 }

\usepackage{microtype}
\usepackage{calrsfs}
\DeclareMathAlphabet{\pazocal}{OMS}{zplm}{m}{n}
\usepackage[cal=rsfso,calscaled=.96]{mathalfa}
\DeclareMathAlphabet{\mathpzc}{OT1}{pzc}{m}{it}
\DeclareMathAlphabet{\mathcalligra}{T1}{calligra}{m}{n}

\def\uu{{\bm u}}
\def\BB{{\textbf {B}}}
\def\bb{{\bm b}}

\def\jm{{\bm {j}}}
\def\JJ{{\textbf {J}}}

\def\xx{{\bm x}}
\def\Nabla{{\pmbmath {\nabla}}}
\def\bomega{{\pmbmath {\omega}}}
\def\be{\begin{equation}}
\def\ee{\end{equation}}
\def\ba{\begin{eqnarray}}
\def\ea{\end{eqnarray}}

\newcommand{\modify}{\color{black}}

\def\bb{{\mathbfcal B}}

\def\bbb{{\mathbf b}}
\usepackage{geometry}
\begin{document}

\title{{{Scale to scale energy transfer rate in compressible two-fluid plasma turbulence}}}

\author{Supratik Banerjee}
\email{sbanerjee@iitk.ac.in}
\affiliation{Department of Physics, Indian Institute of Technology Kanpur, Kalyanpur 208016, Uttar Pradesh, India}
\author{Nahuel Andr{\'e}s}
\email{nandres@iafe.uba.ar}
\affiliation{Institute of Astronomy and Space Physics, Ciudad Universitaria, Buenos Aires, Argentina}
\affiliation{Physics Department, University of Buenos Aires, Ciudad Universitaria, Buenos Aires, Argentina}

\date{\today}

\begin{abstract}
  {We derive the exact relation for the energy transfer in three-dimensional compressible two-fluid plasma turbulence. In the long-time limit, we obtain an exact law which expresses the scale-to-scale average energy flux rate in terms of two point increments of the fluid variables of each species, electric and magnetic field and current density, and puts a strong constraint on the turbulent dynamics. The incompressible single fluid and two-fluid limits and the compressible single fluid limit are recovered under appropriate assumption. {\modify In the single fluid limits, analyses are done with and without neglecting the electron mass thereby making the exact relation suitable for a broader range of application.} In the compressible two fluid regime, the total energy flux rate, unlike the single fluid case, is found to be unaltered by the presence of a background magnetic field. The exact relation provides a way to test whether a range of scales in a plasma is inertial or dissipative and is essential to understand the nonlinear nature of both space and dilute astrophysical plasmas.}
\end{abstract}

\maketitle

\section{Introduction}
Turbulence is a highly nonlinear and complex phenomenon which is ubiquitous in nature. Starting from the tap water, turbulence is found in almost all natural fluids including astrophysical plasmas. In comparison with neutral fluids, plasma turbulence is more difficult to handle due to the presence of more than one characteristic scale (for instance, separately for ions and electrons), which involves different nonlinear dynamic regimes and probable dissipation scales e.g., ion and electron inertial length scales \citep{Bruno05}. Beyond the hydrodynamic length and time scales, every species population (ions, electrons, neutral atoms) of a plasma can be modelled as a separate fluid whence a multi-fluid model can be appropriate to describe the entire plasma. If the plasma is highly ionized such that at every point the ionic and electronic charge densities are nearly equal, a single-fluid model, often called the Extended Magnetohydrodynamics (ExMHD), can be adopted.  The ExMHD can be reduced to the popular Hall-MHD (HMHD) model when the electron mass is neglected with respect to the ion mass. Finally, if one is interested in the fluctuations of sufficiently large length scales (much greater than the ion inertial length or ion gyroradius) and time scales (much greater than the ion gyroperiod), the ordinary Magnetohydrodynamics (MHD) limit can be recovered. Due to its remarkable simplicity and interesting properties, this model has extensively been used to investigate plasma turbulence using analytical, numerical and observational studies \citep{P1998a,M1982,F1995}. 

Whilst the initial understanding of energy transfer across different length scales of incompressible hydrodynamic (IHD) turbulence was principally phenomenological in nature, the pioneering works of \citet{vkh1938} and Kolmogorov built the foundation of rigorous analytical exact relations for IHD turbulence. Assuming statistical stationarity, homogeneity and isotropy, in the limit of infinitely large Reynolds number, \citet{K1941a,K1941b} (hereafter K41) derived an expression for the average energy flux rate $\varepsilon$ in terms of third-order velocity structure functions. This expression is only valid in the inertial range, i.e., the range of scales which are free from both the large-scale forcing and the small-scale dissipation dynamics. This first exact relation, also known as the {\it 4/5 law}, represents one of the cornerstones of turbulence theories using two-point statistics \citep{F1995} and can be written as,
\begin{equation}
\left\langle \delta u_\ell^3 \right\rangle = - \frac{4}{5} \varepsilon \ell,
\end{equation}  
where $u_\ell$ is the fluid velocity component along the increment direction $\boldsymbol\ell$, $\delta \xi$ denotes the difference of a physical quantity $\xi$ between two arbitrary points $\xx$ and $\xx'$ ($\equiv \ \xx + \boldsymbol\ell$) and $\left\langle \cdot \right\rangle$ stands for statistical average. Later, \citet{MY1975} (hereafter MY75) proposed a vectorial form of the above relation without assuming isotropy as,
\begin{equation}
\boldsymbol\nabla_{\ell}\cdot \left\langle \delta u^2 \delta \uu \right\rangle = - 4 \varepsilon{.} \label{vec1}
\end{equation}
These exact results are crucial as they provide an accurate measure of the energy dissipation rate, and therefore, the heating rate of a system by the process of a turbulent cascade \citep{B2016c}. Following both K41 and MY75 formalism, \citet{P1998a,P1998b} derived exact relations for incompressible magnetohydrodynamics (IMHD) turbulence both in terms of (i) fluid velocity and magnetic fields and (ii) the Els\"asser variables, respectively. Other exact laws for incompressible turbulence were derived for MHD, HMHD and ExMHD (with and without electron inertia) turbulence \citep{P2003,Ga2008, Ga2008e, Meyrand10} describing the scale-to-scale transfer of total energy or other inviscid invariants (cross helicity, magnetic helicity, etc.) deep inside the inertial range. However, \citet{Ga2008, Ga2008e} reported that unlike the case of ordinary IHD and IMHD, the average energy flux rate in incompressible HMHD (IHMHD) and electron MHD (ExMHD without electron inertia) turbulence cannot be expressed purely in terms of two-point increments. This issue was successfully dealt using an alternative formulation presented by \citet{B2017} (BG17 hereafter), which expresses the energy cascade rate $\varepsilon$ entirely in terms of two-point increments for IHD, IMHD, IHMHD and electron MHD turbulence without any assumption of isotropy. Very recently, this alternative form has been verified using direct numerical simulation \citep{AndresBan, Ferrand2019}.

Seventy years after K41, a number of exact relations were derived for compressible hydrodynamic (CHD) turbulence using both isothermal and polytropic closures \citep{Ga2011,B2014}. Following the MY75 formalism, the general form of compressible exact relations may be written as,
\begin{equation}
- 4 \varepsilon = \boldsymbol{\nabla}_\ell \cdot {\pazocal F} + {\pazocal S} , 
\end{equation}
where only the so-called flux term ${\pazocal F}$ survives for incompressible turbulence. The new source term ${\pazocal S}$, being proportional to the local velocity divergence (i.e., $\boldsymbol{\nabla}\cdot{\bf u}$), vanishes for the flow with constant density. Similar relations for compressible MHD (CMHD) and Hall-MHD (CHMHD) turbulence were also derived using this methodology \citep{B2013, A2017b, A2018}. Recently, for isothermal CHD turbulence, \citet{B2017k} showed that unlike the kinetic energy correlation function, the thermodynamic energy correlator should be multiplied by a factor 1/2 to make it consistent with the spectral mode energy equipartition in the acoustic limit. This modification is found to rule out the role of the correlation between the velocity and the pressure dilatation in the process of energy cascade \citep{A2011}. However, \citet{A2017b} pointed out that flux-source formulation contains some terms which are neither a pure flux term nor as a source term. Again, there are certain terms which could be cast as both flux and source terms. It is only very recently \citep{B2018} that an exact relation has been derived for isothermal and self gravitating CMHD turbulence, generalizing the BG17 findings and also modifying the thermodynamic energy correlator according to linear mode energy equipartition. The final relation is found to be remarkably compact with respect to the previous compressible exact relations and can be easily extended for rotational or HMHD turbulence.

Similar to incompressible turbulence, exact relations are found to be extremely useful for understanding various aspects of compressible turbulence in space and astrophysical fluids. They have been applied to \textit{in-situ} observations in the fast and slow solar winds and also in the terrestrial magnetosheath to investigate the role of density fluctuations in the turbulence dynamics at the MHD scales and the efficient heating of the system by turbulent energy cascade \citep{B2016c,H2017a,H2017b,A2019PRL}. Moreover, the exact laws for isothermal turbulence (both CHD and CMHD) have been validated using numerical simulations \citep[see, e.g.][]{K2013,A2018b}. 

A considerable number of incompressible and compressible exact scaling relations have been derived in both neutral fluid and MHD turbulence. However, the MHD model constitutes only a very simplistic single fluid model of plasma turbulence, which is only applicable for length scales greater than the ion inertial scales. The Hall and electron MHD models, on the contrary, allow to probe partially into the sub-ion scales. Nevertheless, all those models, being the single fluid models of plasma, assume local charge neutrality (equal ion and electron number densities at every point of the flow field), which could not be the case for several natural plasma systems, especially where the plasma is weakly or partially ionized (\eg cold molecular clouds, protoplanetary disks etc.). For those contexts, it is more appropriate to use the two-fluid (TF) description where the plasma is supposed to consist of two separate fluids of electrons and singly charged ions (for the sake of simplicity, we can neglect the neutral fluids for instance and can include without much problem). The local charge neutrality is no longer valid and each fluid has its individual characteristics. Following both the original von K\'arm\'an-Howarth and BG17 methods, the first exact relations for incompressible TF turbulence were derived recently by \citet{A2016b, A2016c} for the total energy (ionic plus electronic kinetic energy and magnetic energy) and for the generalized ion and electron helicity. However, those exact relations were derived assuming quasi-neutrality and hence represent only a very limited subset of incompressible two-fluid turbulence. 

In the current paper, following the principles of the BG17 formulation, we derive an exact relation for three dimensional (3D), homogeneous and compressible two-fluid (CTF) plasma turbulence. For both ion and electron fluids, we use polytropic closures (with different polytropic indices), which are of interest for both laboratory and astrophysical plasmas. The paper is organized as follows, in Sec.\ref{eqs} we present the basic equations for a polytropic CTF plasma. In Sec. \ref{energy} we  demonstrate the conservation of the total energy in the CTF system, while in Sec. \ref{exactlaw} we present a detailed definition of the energy correlators and the derivation of the exact relation. Finally, in Sec. \ref{discussion} and \ref{conclusions}, we investigate different interesting limits and summarize our main findings along with their potential implications in space and astrophysical plasma turbulence. 


\section{Compressible Two-Fluid Model}

\subsection{{Set of equations}}\label{eqs}

In this paper, we consider a CTF plasma which is composed of a singly-charged ionic and an electronic fluid both satisfying polytropic closures with different polytropic indices. Since in general a TF plasma is not locally quasi-neutral, we assume $n_i\neq n_e$, where $n_{i,e}$ denotes the number density for ions and electrons, respectively. In addition, we assume stationary large-scale forcings in the ion and electron momentum equations. Furthermore, for the sake of simplicity, we assume such a flow regime where the production and the loss rates of each species are roughly equal and {\modify each species fluid is undergoing polytropic equation of state with species specific polytropic index $\gamma_s$.} Therefore, the full set of dynamical equations are given by,
\begin{widetext}
\begin{align}
\p_t \rho_s + \pmbmath{\nabla} \cdot \jm_s  &= 0,  \label{continuity}\;\;\;\;\;\;\;\; \\
\partial_t \jm_s + \pmbmath{\nabla} \cdot (\jm_s \otimes \uu_s) &= - \pmbmath{\nabla} p_s +  \rho_{cs} ( {\bf E} +  \uu_s \times \BB ) +  {\bm d_s} + {\bm f_s}, \label{momentum} \\
\p_t \BB &= - \nabla \times {\bf E} , \label{induction} \\
\p_t {\bf E}  &= \frac{1}{\mu_0 \epsilon_0}\pmbmath{\nabla}  \times \BB  - \frac{1}{\epsilon_0} {\bf J}  =  \frac{1}{\mu_0 \epsilon_0}\pmbmath{\nabla}  \times \BB  - \frac{1}{\epsilon_0} \sum_{s} \rho_{cs} \uu_s{,} \, \, \, \, \text{and} \\
\color{black} {p_{s}} &\color{black}{= K_{s} \rho_{s}^{\gamma_{s}}}. 
\label{hd1}
\end{align}
\end{widetext}
where the index $s$ denotes the individual species of ions ($i$) and electrons ($e$). $\rho_s$, $\rho_{cs}$, $\uu_s$, $\jm_s \equiv \rho_s \uu_s$ and $p_s$ denote the mass density, the electric charge density, the fluid velocity, the mass current and the fluid pressure for each species $s$, respectively.  In the above equations, ${\bf E}$ is the electric field, ${\bf B}$ the magnetic field, ${\bf J \left(\equiv \sum_{s} \rho_{cs} \uu_s\ \right)}$ the electric current and $\epsilon_0$ and $\mu_0$ are the free space permitivity and permeability, respectively.  Finally, ${\bm f}_s$ is the stationary large-scale forcings and ${\bm d_i} = - {\bm d_e}=  \nu \rho_r (\uu_e - \uu_i )$ denotes the momentum exchange between the two species (with $\rho_r = {\rho_i \rho_e}(\rho_i + \rho_e)^{-1}$) \citep{Bittencourt} {\modify and $K_s$ is a constant of proportionality. Note that, in the current study, for the sake of simplicity, forcing is added only in the momentum equation and not in the electromagnetic field evolution equations. It means that the energy is injected in the form of kinetic energy. However, electric and magnetic energy component will also be nourished through the interaction of the momentum field ({$\jm$}) and the velocity field ({\bf u}) with the the electric field ({\bf E}) and magnetic field ({\bf B}).}
\vspace{1cm}


\subsection{Total energy conservation}\label{energy}
In this section, we show that the total energy is conserved for a CTF plasma in the absence of any forcing and viscous term. Using the set of Eqs.~\eqref{continuity}-\eqref{hd1} and setting $\bm f_{s}$ and $\bm d_{s}$ to zero, it is straightforward to show that,
\begin{widetext}
\begin{align}
&{\partial_t} \sum_{s} {\rho_s u_s^2}     =   2 \JJ \cdot {\bf E} -  \sum_{s} \left[ \nabla \cdot  \left(u_s^2 \jm_s  \right)   -  \uu_s \cdot \nabla p_s - \gamma_s \jm_s \cdot \nabla e_s \right]  ,  \\
& {\partial_t \sum_{s} \rho_s e_s }  =  \sum_{s}  \left[ \nabla \cdot (\jm_s e_s ) + p_s (\nabla \cdot \uu_s)  \right] \, , \\
&  {\partial_t} \left(\frac{B^2}{ \mu_0} + \epsilon_0 E^2  \right)   = - 2   \left[ \nabla \cdot \left( \frac{{\bf E} \times {\bf B}}{\mu_0}\right) + \JJ \cdot {\bf E} \right] ,  
\end{align}
\end{widetext}
where $e_{s}= {p}_{s}/{(\gamma_{s} - 1) \rho_{s}}$ is the thermodynamic potential energy per unit mass for species $s$. Noting that for a polytropic fluid $\gamma \jm_s \cdot \nabla e_s = \uu_s \cdot \nabla p_s$, we can show that for a CTF plasma the total energy {${\cal E}$} is an inviscid invariant. The total energy density ($\pazocal E$), which is the sum of the densities of kinetic and thermodynamic potential energies of the individual species plus the density of the electromagnetic energy, can be written as,

\begin{equation}
{\pazocal E} = \frac{1}{2} \left(  \sum_s \rho_s u_s^2 + \frac{B^2}{\mu_0} + \epsilon_0  E^2 \right) + \sum_{s} \frac{p_s}{\gamma_s - 1} \label{consE}.
\end{equation}

In the next Sec. \ref{exactlaw}, we derive an exact relation related to the conservation of total energy in the absence of the forcing and the viscous terms.  

\section{Derivation of the exact relation}\label{exactlaw} 

In order to derive the exact relation for CTF turbulence, we define the two-point correlator for the total energy ${\cal E}$. For a neutral polytropic one-fluid, it can be shown that there is equipartition between the average kinetic energy and the average thermodynamic potential energy in the linear modes \citep{Z1995}. This type of equipartition can, in principle, be generalized for a charged fluid as well. Along with this fact, following \citet{B2018}, here we construct the thermodynamic energy correlator which is in accordance with spectral space energy equipartition between the kinetic and potential energies in the acoustic limit {\modify and also is equal to the  thermodynamic energy density in the single point limit (where the primed and unprimed quantities coincide).  To satisfy both the conditions, it turns out that the correlation function will include an appropriate combination of one-point and two-point contributions having the general form,}
\begin{equation}
\frac{(n-1)}{n} \left \langle \rho e \right \rangle + \left \langle \frac{\rho e' + \rho' e}{2 n}\right \rangle{,} \label{correlator}
\end{equation}
and search for the value of $n$ which gives spectral space equipartition of energy in the acoustic limit. For a polytropic fluid, the total energy density of the acoustic mode (${\pazocal E}_a$) is given by \citep{Lan2013},
\begin{equation}
{\pazocal E}_a =   \frac{1}{2} \rho_0 \bm u^2 +  \frac{C^2}{2 \rho_0} {\rho}_1^2 , \label{consEa}
\end{equation}
where $\rho_0$, $\uu$ and $\rho_1$ represent the mean density and the first-order perturbations in velocity and density, respectively. Here, $C= \sqrt{{\gamma P_0}/{\rho_0}}$ denotes the polytropic sound speed. {\modify The two terms on the right hand side} of Eq. \eqref{consEa} correspond to the the acoustic kinetic and the thermodynamic energy densities, respectively. The polytropic two-point energy correlation function can therefore be written as,
\begin{align}
{\pazocal R}_a(\boldsymbol\ell)&=\frac{1}{2}\rho_0\langle\uu\cdot\uu'\rangle+\frac{C^2}{2\rho_0}\langle{\rho_1}{\rho_1}'\rangle \nonumber \\
&= \frac{1}{2}\rho_0\langle\uu\cdot\uu'\rangle+   \frac{\gamma}{2} \frac{P_0}{\rho_0^2}\langle{\rho_1}{\rho_1}'\rangle . \label{corr}
\end{align}
{\modify In the formalism described in Sec. \RomanNumeralCaps{3} of Ref. \citep{B2018}, it was shown, for an isothermal fluid, that the correlator could give the correct value for the thermodynamic energy density in the spectral space if $n=2$ in equation \eqref{correlator}. Generalising the same methodology for a polytropic fluid, we can expand the two-point contribution (it is only that part which can give us the energy density for the thermodynamic part in spectral space) in the general correlation function at acoustic limit (\textit{i.e.} with only first-order perturbations), which gives 
\begin{widetext}
\begin{align}
&\left \langle \frac{\rho e' + \rho' e}{2 n}\right \rangle = \frac{P_0}{ 2n (\gamma -1)} \left\langle \left(1 + \frac{\rho_1}{\rho_0} \right)  \left(1 + \frac{\rho_1'}{\rho_0} \right)^{\gamma - 1} + \left(1 + \frac{\rho_1'}{\rho_0} \right)  \left(1 + \frac{\rho_1}{\rho_0} \right)^{\gamma - 1} \right\rangle \nonumber \\
&\approx \frac{P_0}{ 2n (\gamma -1)} \left\langle \left(1 + \frac{\rho_1}{\rho_0} \right)  \left[1 + \left( \gamma - 1 \right) \frac{\rho_1'}{\rho_0} \right] + \left(1 + \frac{\rho_1'}{\rho_0} \right)  \left[1 +  \left( \gamma - 1 \right) \frac{\rho_1}{\rho_0} \right] \right\rangle 
\end{align}
\end{widetext}
The total two-point contribution of second order of smallness is clearly given by $\frac{P_0}{n \rho_0^2}\langle{\rho_1}{\rho_1}'\rangle$. Equating this contribution to the thermodynamic energy contribution in Eqn.~\eqref{corr}, we have $n= 2/\gamma$}
(for the isothermal case, $\gamma=1$ and one recovers $n=2$ \citep{B2018}). Therefore, we define the two-point symmetric correlator of the total energy for the TF plasma as, ${\pazocal R}(\boldsymbol\ell) = \left\langle {R_\pazocal{E} + R'_\pazocal{E}} \right\rangle/2$ with
\begin{widetext}
\begin{align}
R_{\pazocal E} &\equiv \frac{1}{2} \left[ \sum_{s} \left(  \jm_s  \cdot {\uu}_s'  + \gamma_s  {\rho_s} {e_s}' + (2-\gamma_s) \rho_s e_s \right) + \bb \cdot \bb'  +   {\mathbfcal E} \cdot {\mathbfcal E'} \right]  {,} \\
R_{\pazocal E}' &\equiv \frac{1}{2} \left[ \sum_{s} \left(  \jm_s'  \cdot {\uu}_s  + \gamma_s  {\rho_s'} {e_s} + (2-\gamma_s) \rho_s' e_s' \right) + \bb' \cdot \bb  +   {\mathbfcal E'} \cdot {\mathbfcal E} \right]  {,} \label{ecf}
\end{align} 
where $ \bb = \BB/ {\sqrt \mu_0}$ and $ \mathbfcal E = {\sqrt \epsilon_0} \bf E$. Using Eqs.~\eqref{continuity}-\eqref{hd1} (including the forcing and dissipation terms) and the two-point statistics for homogeneous turbulence, we obtain,

\begin{align}
\p_t \left\langle \jm_s \cdot \uu_s' \right\rangle &=  \left\langle  \jm_s \cdot \left[ - \Nabla' \left( \gamma_s e_s'  + \frac{u_s'^2}{2}\right) + \left( \uu_s' \times \pmbmath{\omega}_s' \right) + \frac{q_s}{m_s} \left({\bf E}' + \uu_s' \times {\bf B}' \right)  +  \frac{1}{\rho_s'} ({\bm d'_{s}} + {\bm f_s'}) \right]  \right\rangle  \label{e1}  \\
&+ \left\langle  \uu_s' \cdot \left[  - \uu_s \left( \nabla \cdot \jm_s \right) - \rho_s \nabla \left( \frac{u_s^2}{2} \right) +  \left(\jm_s  \times \pmbmath{\omega}_s \right) - \pmbmath{\nabla} p_s + \rho_{cs} \left({\bf E} + \uu_s \times {\bf B} \right)  +  ({\bm d_s} + {\bm f_s}) \right]  \right\rangle, \nonumber \\
\p_t \left\langle {\bb} \cdot {\bb}' \right\rangle &=  - \frac{1}{\mu_0} \langle  \BB \cdot \left( \nabla' \times {\bf E}'  \right) +   \BB' \cdot \left( \nabla \times {\bf E}  \right)  \rangle \label{e3} {,}  \\
\p_t \left\langle {\mathbfcal E} \cdot {\mathbfcal E}' \right\rangle 
&=  \frac{1}{\mu_0} \langle (\Nabla \times \BB) \cdot \bf E'  + (\Nabla' \times \BB') \cdot \bf E \rangle - \langle \JJ \cdot  \bf E'  + \JJ'  \cdot  \bf E  \rangle \label{e4} {,} \\
\p_t \langle {\rho_s} e'_s \rangle &= \left\langle - \frac{\rho_s}{\gamma_s}  \theta'_s {{C_s}'}^2 + \jm_s \cdot\Nabla'e'_s  - \rho_s \uu'_s \cdot\Nabla'e'_s \right\rangle ,  \label{e5}\\
\p_t \langle \rho_s e_s \rangle &= \p_t \langle \rho'_s e'_s \rangle = - \left\langle p_s \theta_s \right\rangle ,  \label{e6}
\end{align} 
\end{widetext}
where $ \theta_s \equiv \boldsymbol\nabla \cdot \uu_s$. By symmetry, we can easily calculate the $\p_t \langle \jm'_s \cdot \uu_s \rangle$ and $\p_t \langle {\rho'_s} e_s \rangle $ correlators. Finally, with these expressions, the evolution equation of the total correlation function ${\pazocal R}$ can be written as, 
\begin{widetext}
\begin{align}
\p_t {\pazocal R} &= \frac{1}{4} \p_t \left \langle \jm_i \cdot \uu_i' + \jm_e \cdot \uu_e' + \jm_i' \cdot \uu_i + + \jm_e' \cdot \uu_e + 2 \bb \cdot \bb' - 2 \mathbfcal E \cdot \mathbfcal E'  \right.  \\
&+ \left. \gamma_i \rho_i e'_i + (2 - \gamma_i) \rho_i e_i + \gamma_e \rho_e e'_e + (2 - \gamma_e) \rho_e e_e + \gamma_i \rho'_i e_i + (2- \gamma_i)  \rho'_i e'_i + \gamma_e \rho'_e e_e + (2- \gamma_e)  \rho'_e e'_e  \right\rangle \nonumber  {.}
\end{align} 
Using Eqs.~\eqref{e1}-\eqref{e6}, we obtain

\begin{align}
\p_t {\pazocal R} &= \frac{1}{4} \left\langle - (\uu'_i \cdot \uu_i) (\nabla \cdot \jm_i) - \rho_i \nabla \left(\frac{u_i^2}{2} \right) \cdot \uu'_i + \uu'_i \cdot (\jm_i \times {\bomega}_i) - \uu'_i \cdot \nabla p_i + \rho_{ci}  \uu'_i \cdot \left({\bf E} + \uu_i \times {\bf B} \right)  + \uu'_i \cdot ({\bm d_i} + {\bm f_i}) \right.  \nonumber \\
& \left. - \jm_i  \cdot \nabla' \left( \gamma_i e'_i + \frac{{u'_i}^2}{2} \right) + \jm_i \cdot ( \uu'_i \times \bomega'_i ) + \frac{e \jm_i}{m_i} \cdot \left({\bf E}' + \uu_i' \times {\bf B}' \right) + \frac{\jm_i}{\rho'_i} \cdot ({\bm d'_i} + {\bm f'_i})  \right. \nonumber \\
& \left. - (\uu'_e \cdot \uu_e) (\nabla \cdot \jm_e) - \rho_e \nabla \left(\frac{u_e^2}{2} \right) \cdot \uu'_e + \uu'_e \cdot (\jm_e \times {\bomega}_e) - \uu'_e \cdot \nabla p_e + \rho_{ce}  \uu'_e \cdot \left({\bf E} + \uu_e \times {\bf B} \right)  + \uu'_e \cdot ({\bm d_e} + {\bm f_e}) \right.  \nonumber \\
& \left. - \jm_e  \cdot \nabla' \left( \gamma_e e'_e + \frac{{u'_e}^2}{2} \right) + \jm_e \cdot ( \uu'_e \times \bomega'_e) - \frac{e \jm_e}{m_e} \cdot \left({\bf E}' + \uu_e' \times {\bf B}' \right) + \frac{\jm_e}{\rho'_e} \cdot ({\bm d'_e} + {\bm f'_e})  \right. \nonumber \\
& \left. - (\uu_i \cdot \uu'_i) (\nabla' \cdot \jm'_i) - \rho'_i \nabla' \left(\frac{{u'_i}^2}{2} \right) \cdot \uu_i + \uu_i \cdot (\jm'_i \times {\bomega}'_i) - \uu_i \cdot \nabla' p'_i + \rho'_{ci}  \uu_i \cdot \left({\bf E'} + \uu'_i \times {\bf B'} \right)  + \uu_i \cdot ({\bm d'_i} + {\bm f'_i}) \right.  \nonumber \\
& \left. - \jm'_i  \cdot \nabla \left( \gamma_i e_i + \frac{{u_i}^2}{2} \right) + \jm'_i \cdot ( \uu_i \times \bomega_i ) + \frac{e \jm'_i}{m_i} \cdot \left({\bf E} + \uu_i \times {\bf B} \right) + \frac{\jm'_i}{\rho_i} \cdot ({\bm d_i} + {\bm f_i})  \right. \nonumber \\
& \left. - (\uu_e \cdot \uu'_e) (\nabla' \cdot \jm'_e) - \rho'_e \nabla' \left(\frac{{u'}_e^2}{2} \right) \cdot \uu_e + \uu_e \cdot (\jm'_e \times {\bomega'}_e) - \uu_e \cdot \nabla' p'_e + \rho'_{ce}  \uu_e \cdot \left({\bf E'} + \uu'_e \times {\bf B'} \right)  + \uu_e \cdot ({\bm d'_e} + {\bm f'_e}) \right.  \nonumber \\
& \left. - \jm'_e  \cdot \nabla \left( \gamma_e e_e + \frac{{u_e}^2}{2} \right) + \jm'_e \cdot ( \uu_e \times \bomega_e) - \frac{e \jm'_e}{m_e} \cdot \left({\bf E} + \uu_e \times {\bf B}  \right) + \frac{\jm'_e}{\rho_e} \cdot ({\bm d_e} + {\bm f_e})  \right. \nonumber \\
& \left. - \frac{2}{\mu_0}  \left[ \BB \cdot \left( \nabla' \times {\bf E}'  \right) +   \BB' \cdot \left( \nabla \times {\bf E}  \right) -  (\Nabla \times \BB) \cdot \bf E'  - (\Nabla' \times \BB') \cdot \bf E \right] - 2 \left( \JJ \cdot  \bf E'  + \JJ'  \cdot  \bf E \right)  \right.  \nonumber \\
& \left. - \gamma_i  \left[ \frac{ p'_i}{\rho'_i} \theta'_i \rho_i - \rho_i \uu_i \cdot \nabla' e'_i + \rho_i \uu'_i  \cdot \nabla' e'_i \right]  - \gamma_e  \left[ \frac{ p'_e}{\rho'_e} \theta'_e \rho_e - \rho_e \uu_e \cdot \nabla' e'_e + \rho_e \uu'_e \cdot  \nabla' e'_e \right] \vphantom{\frac{1}{2}} \right. \nonumber \\
&\left.  - \gamma_i  \left[ \frac{ p_i}{\rho_i} \theta_i \rho'_i - \rho'_i \uu'_i \cdot \nabla e_i + \rho'_i \uu_i  \cdot \nabla e_i \right]  - \gamma_e  \left[ \frac{ p_e}{\rho_e} \theta_e \rho'_e - \rho'_e \uu'_e \cdot \nabla e_e + \rho'_e \uu_e \cdot  \nabla e_e \right]  \vphantom{\frac{1}{2}}  \right. \nonumber\\
&\left. - \left(2-\gamma_i \right) p_i \theta_i  - \left(2-\gamma_e \right) p_e \theta_e  - \left(2-\gamma_i \right) p'_i \theta'_i - \left(2-\gamma_e \right) p'_e \theta'_e \vphantom{\frac{1}{2}} \right\rangle{.} \label{hola}
\end{align}
Using statistical homogeneity, we can show,
\begin{align}
 &\left \langle \left[ \BB \cdot \left( \Nabla' \times {\bf E}'  \right) +   \BB' \cdot \left( \Nabla \times {\bf E}  \right) -  (\Nabla \times \BB) \cdot \bf E'  - (\Nabla' \times \BB') \cdot \bf E \right] \right \rangle 
= \\ 
 &\left \langle  \Nabla' \cdot \left(  {\bf E}'  \times \BB  \right) 
+   \Nabla \cdot \left( {\bf E} \times {\BB}'  \right) \right.  -  \left. \Nabla \cdot (\BB \times {\bf E}') - \Nabla' \cdot (\BB' \times \bf E ) \right \rangle  
= \\ 
 &\Nabla_{\bf \ell} \cdot  \left \langle \left(  {\bf E}'  \times \BB  \right) -  \left( {\bf E} \times {\BB}'  \right) +   (\BB \times {\bf E}') -  (\BB' \times \bf E )  \right \rangle = 0.
\end{align}
Again Eq.~\eqref{hola} yields,

\begin{align}
4 \p_t {\pazocal R} = & \left\langle \delta \jm_i   \cdot   \delta \left[ \left( \uu_i \cdot \nabla \right) \uu_i \right] + \delta \uu_i \cdot  \delta \left[ \nabla \cdot \left( \jm_i \otimes \uu_i \right) \right] + \delta \jm_e   \cdot   \delta \left[ \left( \uu_e \cdot \nabla \right) \uu_e \right] + \delta \uu_e \cdot  \delta \left[ \nabla \cdot \left( \jm_e \otimes \uu_e \right) \right] \right \rangle \nonumber \\
&+ \gamma_i \left\langle \delta \rho_i \delta \left[ \frac{p_i \theta_i }{\rho_i} + \left( \uu_i \cdot \nabla \right) e_i \right] + \delta \uu_i \cdot \delta \left[ \rho_i \nabla e_i \right] \right\rangle  + \gamma_e \left\langle \delta \rho_e \delta \left[ \frac{p_e \theta_e }{\rho_e} + \left( \uu_e \cdot \nabla \right) e_e \right] + \delta \uu_e \cdot \delta \left[ \rho_e \nabla e_e \right] \right\rangle  \nonumber  \\
&-  \left\langle  \delta (\rho_{ci} \uu_i) \cdot \delta (\uu_i \times \BB) + \delta \uu_i \cdot \delta (\rho_{ci} \uu_i \times \BB) + \delta (\rho_{ce} \uu_e) \cdot \delta (\uu_e \times \BB) + \delta \uu_e \cdot \delta (\rho_{ce} \uu_e \times \BB)  \right\rangle \nonumber \\
&+ \left\langle \delta { \JJ} \cdot \delta {\bf E} - \delta \uu_i \cdot \delta (\rho_{ci} {\bf E}) - \delta \uu_e \cdot \delta (\rho_{ce} {\bf E})  \right\rangle + 4 {\mathbfcal D} + 4 {\mathbfcal F} {.} \label{finalexact}
\end{align}
Finally, we assume a stationary state for which the left hand term of the above equation vanishes and we restrict ourselves to the length scales far away of the dissipation length scales whence we can neglect ${\mathbfcal D}$ . As a result, we can identify ${\mathbfcal F} = \varepsilon$ and the final exact relation becomes, 
\begin{align}
-4 \varepsilon = & \left\langle \delta \jm_i   \cdot   \delta \left[ \left( \uu_i \cdot \nabla \right) \uu_i \right] + \delta \uu_i \cdot  \delta \left[ \nabla \cdot \left( \jm_i \otimes \uu_i \right) \right] + \delta \jm_e   \cdot   \delta \left[ \left( \uu_e \cdot \nabla \right) \uu_e \right] + \delta \uu_e \cdot  \delta \left[ \nabla \cdot \left( \jm_e \otimes \uu_e \right) \right] \right \rangle \nonumber \\
&+ \gamma_i \left\langle \delta \rho_i \delta \left[ \frac{p_i \theta_i }{\rho_i} + \left( \uu_i \cdot \nabla \right) e_i \right] + \delta \uu_i \cdot \delta \left[ \rho_i \nabla e_i \right] \right\rangle  + \gamma_e \left\langle \delta \rho_e \delta \left[ \frac{p_e \theta_e }{\rho_e} + \left( \uu_e \cdot \nabla \right) e_e \right] + \delta \uu_e \cdot \delta \left[ \rho_e \nabla e_e \right] \right\rangle  \nonumber  \\
&-  \left\langle  \delta (\rho_{ci} \uu_i) \cdot \delta (\uu_i \times \BB) + \delta \uu_i \cdot \delta (\rho_{ci} \uu_i \times \BB) + \delta (\rho_{ce} \uu_e) \cdot \delta (\uu_e \times \BB) + \delta \uu_e \cdot \delta (\rho_{ce} \uu_e \times \BB)  \right\rangle \nonumber \\
&+ \left\langle \delta { \JJ} \cdot \delta {\bf E} - \delta \uu_i \cdot \delta (\rho_{ci} {\bf E}) - \delta \uu_e \cdot \delta (\rho_{ce} {\bf E})  \right\rangle {.} \label{finalexact}
\end{align}
\end{widetext}
Equation \eqref{finalexact} represents an exact law for energy transfer rate in homogeneous, polytropic TF plasma turbulence. It is worth mentioning that the final exact relation Eq.~\eqref{finalexact} is written only as a function of two-point increments and it is valid only in the inertial range (i.e., for length scales far away from the injection and dissipative scales). {\modify As discussed in \citep{B2018}, unlike incompressible turbulence, the net turbulent contribution in the energy flux rate cannot simply be expressed as a departure from the aligned states (generalized Beltrami flows). In the incompressible limit of two-fluid turbulence (see the next section), one can express $\varepsilon$ as a departure from different aligned states.}  In the next Sec. \ref{discussion}, we discuss various important features of Eq.~\eqref{finalexact}.

\section{Discussion}\label{discussion}

\subsection{Incompressible TF limit}
In the incompressible limit, both the ion and electron number densities, i.e., $n_i$ and $n_e$ (and hence the mass $\rho_{e,i}$ and charge densities $\rho_{ce,ci}$), are constants in time and space. However, they are not necessarily equal. Using this assumption and statistical homogeneity, one can show for each species $s$, 
\begin{widetext}
\begin{equation}
\left\langle \delta \jm_s   \cdot   \delta \left[ \left( \uu_s \cdot \nabla \right) \uu_s \right] + \delta \uu_s \cdot  \delta \left[ \nabla \cdot \left( \jm_s \otimes \uu_s \right) \right] \right\rangle = 2 \rho_s \left\langle \delta \uu_s   \cdot   \delta \left[ \left( \uu_s \cdot \nabla \right) \uu_s \right] \right\rangle = - 2 \rho_s \left\langle \delta \uu_s   \cdot   \delta \left( \uu_s \times \bomega_s \right) \right\rangle  \nonumber
\end{equation}
where $\nabla \cdot \uu_s=0$. The thermodynamic energy terms also vanish for incompressible fluid. The, Eq.~\eqref{finalexact} is therefore simplified as,

\begin{equation}
2 \varepsilon =    \left\langle  \rho_{i0}  \delta \uu_i   \cdot   \delta \left( \uu_i \times \bomega_i \right)  +  \rho_{e0}  \delta \uu_e   \cdot   \delta \left( \uu_e \times \bomega_e \right)  
+  \rho_{ci0} \delta  \uu_i \cdot \delta (\uu_i \times \BB) +  \rho_{ce0} \delta  \uu_e \cdot \delta (\uu_e \times \BB) \right\rangle.  \label{incompressible}
\end{equation}
\end{widetext}
where  $\rho_{e0}$ and $\rho_{i0}$ are the mean mass densities and $\rho_{ce0}$ and $\rho_{ci0}$ are the mean charge densities for electrons and ions, respectively.
{\modify As expected for incompressible turbulence, here, the turbulent energy flux rate $\varepsilon$ is written in terms of the  departure from aligned states which can be obtained as minimum energy states along with the constraints of the conservations of generalized helicities of both ions and electrons \citep{S1997}. The two relaxed states in a barotropic or even in an incompressible two-fluid plasma are represented by the two following alignments:
$
(m_{i,e} \bomega_{i,e} + q_{i,e} {\bf B}) \parallel n_{i,e} q_{i,e} \uu_{i,e}
$. The above equation \eqref{incompressible} evidently shows that $\varepsilon$ reduces to zero when both the alignments occur. In addition if the electron mass is neglected, then the corresponding alignment will simply be an alignment between electron fluid velocity and the magnetic field thereby leading to a  magnetic force-free motion of the electrons.} Interestingly, in the above equation, the terms associated directly with the electric field {\bf E}, vanish under the incompressibility assumption. It is worth mentioning that Eq.~\eqref{incompressible} is valid only when the ion and electron density fluctuations are equal to zero. However, its mean values ($\rho_{e0}$ and $\rho_{i0}$) are not necessarily equal. For a TF plasma with very strong ionisation, the charge quasi-neutrality assumption, i.e., $n_i=n_e=n_0$ can be satisfied and Eq.~\eqref{incompressible} then reduces to, 	
\begin{widetext}
\begin{align}
2 \varepsilon &=    n_0 \left\langle  m_i  \delta \uu_i   \cdot   \delta \left( \uu_i \times \bomega_i \right)  +  m_e  \delta \uu_e   \cdot   \delta \left( \uu_e \times \bomega_e \right)  
+  e \delta  \uu_i \cdot \delta (\uu_i \times \BB) -  e \delta  \uu_e \cdot \delta (\uu_e \times \BB) \right\rangle  \label{reduction} \\   
&= n_0 M \left\langle   \delta \uu_i   \cdot   \delta \left[ \uu_i \times \left( \frac{m_i}{M} \bomega_i + \frac{e}{M} \BB \right) \right]  +   \delta \uu_e  \cdot   \delta \left[ \uu_e \times \left( \frac{m_e}{M} \bomega_e - \frac{e}{M} \BB \right) \right]  \right\rangle	\nonumber \\
&= 	n_0 M \left\langle   \delta \uu_i   \cdot   \delta \left[ \uu_i \times \left( \frac{e}{M} \sqrt{\mu_0 n_0 M} {\bf b} + \frac{m_i}{M} \bomega_i\right) \right]  -   \delta \uu_e  \cdot   \delta \left[ \uu_e \times \left( \frac{e}{M} \sqrt{\mu_0 n_0 M} {\bf b} - \frac{m_e}{M} \bomega_e  \right) \right]  \right\rangle \nonumber \\ 
&= 	n_0 M \left\langle   \delta \uu_i   \cdot   \delta \left[ \uu_i \times \left( \frac{e \sqrt{\mu_0 n_0}}{\sqrt{M}}  {\bf b} + (1 - \mu) \bomega_i\right) \right]  -   \delta \uu_e  \cdot   \delta \left[ \uu_e \times \left( \frac{e \sqrt{\mu_0 n_0}}{\sqrt{M}}  {\bf b} - \mu \bomega_e  \right) \right]  \right\rangle \nonumber \\,  
&= 	n_0 M \left\langle   \delta \uu_i   \cdot   \delta \left[ \uu_i \times \left( \frac{\bf b}{\lambda} + (1 - \mu) \bomega_i\right) \right]  -   \delta \uu_e  \cdot   \delta \left[ \uu_e \times \left( \frac{\bf b}{\lambda} - \mu \bomega_e  \right) \right]  \right\rangle \nonumber  \\
&= 	\frac{n_0 M}{\lambda} \left\langle   \delta \uu_i   \cdot   \delta \left[ \uu_i \times \left( {\bf b} + \lambda (1 - \mu) \bomega_i\right) \right]  -   \delta \uu_e  \cdot   \delta \left[ \uu_e \times \left( {\bf b} - \lambda \mu \bomega_e  \right) \right]  \right\rangle \nonumber
\end{align}
\end{widetext}
which has previously been derived by \citet{A2016c}. In the above expressions, $M= {m_i+ m_e}$, $\mu = {m_e}/{M}$, {\modify $ {\bf b} = \frac{\BB}{\sqrt{\mu_0 n_0 M}}$} and $\lambda=  {\sqrt{M}}/\left({e \sqrt{\mu_0 n_0}}\right)$. Similar to the IHD and IMHD turbulence, the turbulent energy flux in a two-fluid plasma can also be attributed to the departure of Beltrami type of alignments for both ionic and electronic fluids. For aligned cases $i.e.$, for $\uu_{i,e} \parallel \bomega_{i,e}$  $\uu_{i,e} \parallel \BB$, the contribution of the corresponding term to $\varepsilon$ vanishes \citep[see also,][]{B2017}. 

\subsection{Compressible single fluid limits}

{\modify
\paragraph{\bf Without neglecting electron mass:} 
In the single fluid limit, at every point  $n_i = n_e= n$, however, $n$ is not necessarily constant. Now from the definitions of single fluid variables, we can write,
\begin{align}
\uu&= \frac{n_i m_i \uu_i + n_e m_e \uu_e}{n_i m_i + n_e m_e}= \frac{m_i \uu_i + m_e \uu_e}{m_i + m_e} \\
\JJ &= e (n_i \uu_i - n_e \uu_e) = n e (\uu_i - \uu_e)
\end{align}
Solving for $\uu_i$ and $\uu_e$, we get 
 \begin{align}
\uu_i &=  \uu + \frac{\alpha}{1 + \alpha} \left( \frac{\JJ}{n e} \right) \, \, \, \, \text{and} \\
\uu_e &=  \uu - \frac{1}{1 + \alpha} \left( \frac{\JJ}{n e} \right),
\end{align} where $\alpha \equiv m_e/m_i = \mu/(1- \mu)$. Here we first derive the single fluid limit without neglecting electron mass and so that $\alpha \ll 1$ but $\alpha \neq 0$. This is important because the electron mass is responsible for magnetic reconnection to occur even in a collisionless plasma \citep{Andres14} and turbulent reconnection is believed to affect the turbulent cascade \citep{Comisso18}.
The value of $\alpha$ depends on the mass of the ion and is maximum $(1/1837= 0.000544)$ for the lightest Hydrogen ions. So, for practical purpose, the effect of electron mass can be considered upto the first order of $\alpha$. This limit will lead to a regime of MHD which is more general than ordinary or Hall MHD and therefore can be called MHD with electron inertia. Note that this limit is nevertheless more restricted than the extended MHD model described in some previous works \citep{Abdelhamid} which did not neglect ${\partial \JJ}/{\partial t}$, $\nabla \cdot \left( \uu \otimes \JJ + \JJ \otimes \uu \right)$ and the electron pressure gradient term in Generalized Ohm's laws. These terms are certainly of theoretical interest. However, for most practical cases, these contributions are always neglected. Using the aforesaid expressions of ion and electron velocities, we can write 
\begin{widetext}
\begin{align}
\left( \uu_i \cdot \nabla \right) \uu_i &= \left( \uu \cdot \nabla \right) \uu  +  \frac{\alpha}{1 + \alpha} \left[ \left( \Jc \cdot \nabla \right) \uu + \left( \uu \cdot \nabla \right) \Jc \right] + \frac{\alpha^2}{\left(1 + \alpha \right)^2} \left( \Jc \cdot \nabla \right) \Jc \\
\left( \uu_e \cdot \nabla \right) \uu_e &= \left( \uu \cdot \nabla \right) \uu  -  \frac{1}{1 + \alpha} \left[ \left( \Jc \cdot \nabla \right) \uu + \left( \uu \cdot \nabla \right) \Jc \right] + \frac{1}{\left(1 + \alpha \right)^2} \left( \Jc \cdot \nabla \right) \Jc 
\end{align}
Using the above expressions, we obtain (keeping the terms upto first order of $\alpha$)
\begin{align}
&\delta \jm_i   \cdot   \delta \left[ \left( \uu_i \cdot \nabla \right) \uu_i \right] +
 + \delta \jm_e   \cdot   \delta \left[ \left( \uu_e \cdot \nabla \right) \uu_e \right] \nonumber \\
 &= \delta \left(\rho \uu \right)  \cdot   \delta \left[ \left( \uu \cdot \nabla \right) \uu \right] + \frac{\alpha}{\left(1 + \alpha \right)^2} \  \delta \left( \rho \Jc \right) \cdot \delta    \left[ \left( \Jc \cdot \nabla \right) \uu + \left( \uu \cdot \nabla \right) \Jc \right] \nonumber \\ 
&  + \frac{\alpha}{\left(1 + \alpha \right)^2} \  \delta \left(\rho \uu \right) \cdot \delta    \left[ \left( \Jc \cdot \nabla \right) \Jc \right] - \frac{\alpha \left( 1 - \alpha \right)}{\left(1 + \alpha \right)^3} \ \delta \left(\rho \Jc \right) \cdot \delta  \left[\left( \Jc  \cdot \nabla \right) \Jc  \right] \\
&\approx  \delta \left(\rho \uu \right)  \cdot   \delta \left[ \left( \uu \cdot \nabla \right) \uu \right] + \alpha  \  \delta \left( \rho \Jc \right) \cdot \delta    \left[ \left( \Jc \cdot \nabla \right) \uu + \left( \uu \cdot \nabla \right) \Jc \right] \nonumber \\ 
&  +\alpha  \  \delta \left[\rho \left( \uu - \Jc \right) \right]  \cdot \delta    \left[ \left( \Jc \cdot \nabla \right) \Jc \right] 
\end{align}

Following the same methodology, we can also write
\begin{align}
\rho_i \uu_i \otimes \uu_i &= \frac{1}{1 + \alpha}\rho  \uu \otimes \uu + \frac{\alpha}{\left(1 + \alpha \right)^2} \left( \rho \Jc \otimes \uu + \rho \uu \otimes \Jc \right) + \frac{\alpha^2}{\left(1 + \alpha \right)^3} \left( \rho \Jc \otimes  \Jc \right), \\
\rho_e \uu_e \otimes \uu_e &= \frac{\alpha}{1 + \alpha}\rho  \uu \otimes \uu -  \frac{\alpha}{\left(1 + \alpha \right)^2} \left( \rho \Jc \otimes \uu + \rho \uu \otimes \Jc \right) + \frac{\alpha}{\left(1 + \alpha \right)^3} \left( \rho \Jc \otimes  \Jc \right)
\end{align}
where by definition, $\rho_i = \frac{\rho}{1 + \alpha} $ and $\rho_e = \frac{\alpha}{1 + \alpha} \rho$. After some steps of straightforward algebra, we get by retaining terms upto first order of $\alpha$,
\begin{align}
&\delta \uu_i \cdot  \delta \left[ \nabla \cdot \left( \jm_i \otimes \uu_i \right) \right] + \delta \uu_e \cdot  \delta \left[ \nabla \cdot \left( \jm_e \otimes \uu_e \right) \right]  \nonumber \\
&= \delta \uu \cdot  \delta \left[ \nabla \cdot \left( \rho \uu \otimes \uu \right) \right] + \frac{\alpha}{\left(1 + \alpha \right)^2} \  \delta \left(\Jc \right) \cdot \delta \left[ \nabla \cdot  \left( \rho \Jc \otimes \uu + \rho \uu \otimes \Jc \right) \right]  \nonumber \\
&+ \frac{\alpha}{\left(1 + \alpha \right)^2} \  \delta \uu \cdot \delta \left[ \nabla \cdot  \left( \rho \Jc \otimes \Jc \right) \right]  - \frac{\alpha \left( 1 - \alpha \right)}{\left(1 + \alpha \right)^3} \ \delta \left(\Jc \right) \cdot \delta  \left[ \nabla \cdot \left( \rho \Jc  \otimes \Jc \right) \right] \\
&\approx \delta \uu \cdot  \delta \left[ \nabla \cdot \left( \rho \uu \otimes \uu \right) \right] + \alpha  \  \delta \left(\Jc \right) \cdot \delta \left[ \nabla \cdot  \left( \rho \Jc \otimes \uu + \rho \uu \otimes \Jc \right) \right] \nonumber  \\
&+ {\alpha} \  \delta \left( \uu - \Jc \right) \cdot \delta \left[ \nabla \cdot  \left( \rho \Jc \otimes \Jc \right) \right] 
\end{align}
For the contribution of the thermodynamic energy part, we can write for any species s, 
\begin{align}
&(i)~\gamma_s \left\langle \delta \rho_s \  \delta \left[ \frac{p_s \theta_s }{\rho_s} + \left( \uu_s \cdot \nabla \right) e_s \right] \right\rangle= \gamma_s m_s^{\gamma_s} \left\langle \delta n  \ \delta \left[  K_s n^{\gamma_s -1} \theta_s  + \left( \uu_s \cdot \nabla \right) \frac{K_s n^{\gamma_s -1}}{\gamma_s -1} \right] \right\rangle, \\
&(ii)~\gamma_s \ \delta \uu_s \cdot \delta \left[ \rho_s \nabla e_s \right] = \gamma_s m_s^{\gamma_s} \ \delta \uu_s \cdot \delta \left[ n \  \nabla \left(\frac{K_s n^{\gamma_s -1}}{\gamma_s -1} \right)  \right].
\end{align}
\end{widetext}
So, just by having an apparent view, one can think that the contribution from the electron fluid is negligibly small as it carries a ratio which is proportional to $\alpha^{\gamma_e}$ ($\gamma_e > 1$) and also all other quantities ($K_s$, $\theta_s$, $\gamma_s$) are supposed to be approximately of the same order . At this point one has to be very careful. In practice, both the ion and electron pressure are of comparative magnitude and in fact with the same number density, electron fluid pressure is greater than ion fluid pressure due to having higher temperature. In terms of the polytropic closure equations, it simply says that $K_e$ is considerably greater than $K_i$ so that both pressures are significant. The total pressure can be written as, 
\begin{widetext}
\begin{equation}
p = p_i + p_e = K_i \rho_i^{\gamma_i} + K_e \alpha^{\gamma_e} \rho_i^{\gamma_e}=  \frac{K_i}{\left( 1 + \alpha \right)^{\gamma_i}} \rho^{\gamma_i} + \frac{K_e  \alpha^{\gamma_e}}{\left( 1 + \alpha \right)^{\gamma_e}}  \rho^{\gamma_e}
\end{equation}
\end{widetext}
So, unlike the component species fluids, the pressure-density closure is not a simple polytropic closure for the resultant single fluid. However, in practice, for most cases, compressible MHD is studied using a simple polytropic closure. Under the condition of comparable ion and electron pressures ($K_i$ and $K_e \alpha^{\gamma_e}$ are comparable), a simple polytropic law for the single fluid is possible when $\gamma_i = \gamma_e = \gamma$, where $\gamma$ is the polytropic index for the single fluid. In that case, one can write $p = K \rho^\gamma$ where $K= {\left(K_i + \alpha^ \gamma  K_e \right)}/{\left(1 + \alpha \right)^\gamma}$. For comparable contribution of ion and electron pressures, we necessarily have $K_i \approx \alpha^\gamma K_e$. Using the one fluid quantities, we can write
\begin{widetext}
\begin{align}
(i)~&\left\langle \gamma_i \delta \rho_i \ \delta \left(\frac{P_i \theta_i}{\rho_i} \right) +  \gamma_e \delta \rho_e \ \delta \left(\frac{P_e \theta_e}{\rho_e} \right) \right\rangle \nonumber \\
&\approx \frac{\gamma}{\left(1 + \alpha\right)^\gamma} \left\langle \delta \rho \ \delta \left[ K_i \rho^{\gamma-1} \left( \theta + \alpha \chi \right) \right] + \alpha^\gamma \ \delta \rho \ \delta \left[ K_e \rho^{\gamma-1} \left( \theta  - \chi + \alpha \chi \right) \right] \right\rangle \nonumber \\
&= \gamma  \left\langle \delta \rho \ \delta \left(\frac{p \theta}{\rho} \right) +  \alpha  \ \delta \rho \ \delta \left(\frac{p \chi}{\rho} \right) - \delta \rho_e \ \delta \left(\frac{p_e \chi}{\rho_e} \right) \right\rangle, \label{internal1} \\
(ii)~&\left\langle \gamma_i \delta \rho_i \ \delta \left[\left(\uu_i \cdot \nabla \right) e_i \right] +  \gamma_e \delta \rho_e \ \delta \left[\left(\uu_e \cdot \nabla \right) e_e \right]\right\rangle \nonumber \\
&\approx \frac{\gamma}{\left(1 + \alpha\right)^\gamma} \left\langle \delta \rho \ \delta \left[ \left( \uu + \alpha {\Jc} \right) \cdot \nabla  \left( \frac{K_i \rho^{\gamma-1}}{\gamma - 1} \right)  \right] + \alpha^\gamma \ \delta \rho \ \delta \left[ \left( \uu + \left( \alpha -1 \right) {\Jc} \right) \cdot \nabla  \left( \frac{K_e \rho^{\gamma-1}}{\gamma - 1} \right)  \right] \right\rangle \nonumber \\
&= \gamma  \left\langle \delta \rho \ \delta \left[ \left( \uu \cdot \nabla \right) e \right] +  \alpha  \ \delta \rho \ \delta \left[ \left( \Jc \cdot \nabla \right) e \right] - \delta \rho_e \ \delta \left[ \left( \Jc \cdot \nabla \right) e_e \right]  \right\rangle, \label{internal2} \\
(iii)~ &\left\langle \gamma_i \ \delta \uu_i \cdot \delta \left[ \rho_i \nabla e_i \right] + \gamma_e \ \delta \uu_e \cdot \delta \left[ \rho_e \nabla e_e \right] \right\rangle \nonumber \\
&\approx \frac{\gamma}{\left(1 + \alpha\right)^\gamma} \left\langle \delta \left( \uu + \alpha {\Jc} \right) \cdot  \ \delta \left[  \rho \ \nabla \left( \frac{K_i \rho^{\gamma-1}}{\gamma - 1} \right)  \right] + \alpha^\gamma \ \delta \left( \uu + \left( \alpha -1 \right) {\Jc} \right) \cdot  \ \delta \left[ \rho  \nabla  \left( \frac{K_e \rho^{\gamma-1}}{\gamma - 1} \right)  \right] \right\rangle \nonumber \\
&=\gamma  \left\langle \delta \uu  \cdot \ \delta  \left( \rho \  \nabla  e \right) +  \alpha  \ \delta \left(\Jc\right) \cdot \ \delta \left( \rho \ \nabla e \right)  - \delta \left(\Jc\right) \cdot \ \delta \left( \rho_e  \nabla \right) e_e   \right\rangle, \label{internal3}
\end{align}
\end{widetext}
where $\chi \equiv \nabla \cdot \left( \Jc \right)$. In the Eqns.~\eqref{internal1}--\eqref{internal3}, one can notice that there are some terms containing electron fluid density, pressure and internal energy. In order to get rid of these terms, we need to introduce a new single fluid variable called the electrokinetic pressure $P^E$ (as discussed in \citep{Bittencourt}) which can be written as
\begin{equation}
p^E= \sum_s \frac{\rho_{cs}}{\rho_s} p_s = e \left( \frac{p_i}{m_i} - \frac{p_e}{m_e} \right).
\end{equation}
Note that, for comparable ion and electron fluid pressures, $P^E$ is effectively negative. Using simple algebra, one can show that 
\begin{equation}
p_e \approx \alpha \left( p - \frac{m_i}{e} p^E \right) , \quad  e_e \approx \frac{ \left(1 + \alpha \right)} {\left(\gamma - 1 \right) } \left( \frac{p}{\rho} - \frac{m_i}{e} \frac{p^E}{\rho}  \right)
\end{equation}
So in principle, one can write the Eqn.~\eqref{finalexact} in terms of single fluid quantities assuming quasi-neutrality and a polytropic closure for the resultant single fluid. However, for the sake of comparison between different contributions, we keep the terms with $\rho_e$, $p_e$ and $e_e$ as they are. 
In the next step, we investigate the reduced form of the total electromagnetic contribution of the exact relation in the single fluid limit. For doing so, we need to use the specific form of Generalized Ohm's law which reduces to the Ohm's law for Hall-MHD in the limit $\alpha \rightarrow 0$. We therefore use the following form \citep{Bittencourt}:
\begin{widetext}
\begin{align}
&n e \left(\frac{1}{m_i} + \frac{1}{m_e} \right) \left( {\bf E} + \uu \times \BB - \frac{\JJ}{\sigma_0} \right) =  \left(\frac{1}{m_e} - \frac{1}{m_i} \right) \left(\JJ \times \BB \right)\\
\implies &  \left( {\bf E} + \uu \times \BB - \frac{\JJ}{\sigma_0} \right) = \left( \frac{ 1 - \alpha}{1 + \alpha } \right)  \frac{\left(\JJ \times \BB \right)}{n e }
\end{align}
where, $\sigma_0 = \frac{n e^2}{ m_e \nu_{ei}}$ is the conductivity of the single fluid. In the ideal limit, where the single fluid has infinite conductivity $(\sigma_0 \rightarrow \infty)$, the resulting Ohm's law can be expressed as
\begin{align}
&\left( {\bf E} + \uu \times \BB  \right) = \left( \frac{ 1 - \alpha}{1 + \alpha } \right)  \frac{\left(\JJ \times \BB \right)}{n e } \, \\ 
\implies &\left( {\bf E} + \uu_i \times \BB  \right) = \left( \frac{ 1 }{1 + \alpha } \right)  \frac{\left(\JJ \times \BB \right)}{n e } \, \\
\implies &\left( {\bf E} + \uu_e \times \BB  \right) = - \left( \frac{ \alpha  }{1 + \alpha } \right)  \frac{\left(\JJ \times \BB \right)}{n e }
\end{align}
\end{widetext}
Now we obtain the total electromagnetic contribution for compressible single fluid limit. For the sake of easier calculation, we start the analysis from the correlator containing the electromagnetic contribution. Using the aforesaid expressions of generalized Ohm's law, the resulting correlation function becomes: 
\begin{widetext}
\begin{align}
&\left\langle \uu' \cdot \left( \JJ \times \BB \right) + \uu \cdot \left( \JJ' \times \BB' \right) + \jm \cdot \left( \frac{\JJ' \times \BB'}{\rho'} \right) +  \jm' \cdot \left( \frac{\JJ \times \BB}{\rho} \right) + 2 \JJ \cdot \left(\uu' \times \BB' \right) + 2 \JJ'  \cdot \left(\uu \times \BB \right) \right. \nonumber \\
&\left. - 2 \left(\frac{1 - \alpha}{1 + \alpha}\right) \left[  \JJ \cdot \left( \frac{\JJ' \times \BB'}{n' e} \right) +  \JJ' \cdot \left( \frac{\JJ \times \BB}{n e} \right) \right] \right\rangle \nonumber \\
&=  \left\langle  \delta \uu \cdot \delta (\BB \times \JJ) + \delta \jm \cdot \delta \left( \frac{\BB \times \JJ}{\rho} \right) + 2  \delta \JJ \cdot \delta \left(\BB \times \uu  \right)  - 2 \left(\frac{1 - \alpha}{1 + \alpha}\right) \delta \JJ \cdot \delta \left( \frac{ \BB \times   \JJ}{n e} \right)  \right\rangle, \label {empart} \\
&\approx \left\langle  \delta \uu \cdot \delta (\BB \times \JJ) + \delta \jm \cdot \delta \left( \frac{\BB \times \JJ}{\rho} \right) + 2 \delta \JJ \cdot \delta \left[ \BB \times  \left(\uu - \frac{\JJ}{n e} \right) \right] + 4 \alpha  \  \delta \JJ \cdot \delta \left( \frac{ \BB \times   \JJ}{n e} \right)  \right\rangle
\end{align}
\end{widetext}
where we have used the vector identity $
\jm \cdot  \left( \frac{\BB \times \JJ}{\rho} \right)= \uu \cdot (\BB \times \JJ) 
$ and in the final expression, we have kept the contribution upto first order of $\alpha$. 
Finally, without neglecting the electron mass but keeping their contribution upto first order of $\alpha$, one can write the exact relation for a TF plasma in the single fluid limit as
\begin{widetext}
\begin{align}
&- 4 \varepsilon= \left\langle \delta \left(\rho \uu \right)  \cdot   \delta \left[ \left( \uu \cdot \nabla \right) \uu \right] + \delta \uu \cdot  \delta \left[ \nabla \cdot \left( \rho \uu \otimes \uu \right) \right] + \alpha  \  \delta \left( \rho \Jc \right) \cdot \delta    \left[ \left( \Jc \cdot \nabla \right) \uu + \left( \uu \cdot \nabla \right) \Jc \right]  \right. \nonumber \\ 
& \left. +\alpha  \  \delta \left[\rho \left( \uu - \Jc \right) \right]  \cdot \delta    \left[ \left( \Jc \cdot \nabla \right) \Jc \right]  + \alpha  \  \delta \left(\Jc \right) \cdot \delta \left[ \nabla \cdot  \left( \rho \Jc \otimes \uu + \rho \uu \otimes \Jc \right) \right]  \right. \nonumber  \\
&\left. + {\alpha} \  \delta \left( \uu - \Jc \right) \cdot \delta \left[ \nabla \cdot  \left( \rho \Jc \otimes \Jc \right) \right] + \gamma \ \delta \rho \ \delta \left(\frac{p \theta}{\rho} \right) +  \alpha  \gamma \ \delta \rho \ \delta \left(\frac{p \chi}{\rho} \right) - \gamma \ \delta \rho_e \ \delta \left(\frac{p_e \chi}{\rho_e} \right) \right.  \nonumber \\
&\left.  +\gamma \ \delta \rho \ \delta \left[ \left( \uu \cdot \nabla \right) e \right] +  \alpha \gamma \  \ \delta \rho \ \delta \left[ \left( \Jc \cdot \nabla \right) e \right] - \gamma \ \delta \rho_e \ \delta \left[ \left( \Jc \cdot \nabla \right) e_e \right] + \gamma \ \delta \uu  \cdot \ \delta  \left( \rho \  \nabla  e \right) \right. \nonumber \\
&\left. +   \alpha \gamma  \ \delta \left(\Jc\right) \cdot \ \delta \left( \rho \ \nabla e \right)  - \gamma \delta \left(\Jc\right) \cdot \ \delta \left( \rho_e  \nabla \right) e_e \right.  \nonumber \\
&\left. + \delta \uu \cdot \delta (\BB \times \JJ) + \delta \jm \cdot \delta \left( \frac{\BB \times \JJ}{\rho} \right) + 2 \delta \JJ \cdot \delta \left[ \BB \times  \left(\uu - \frac{\JJ}{n e} \right) \right] + 4 \alpha  \  \delta \JJ \cdot \delta \left( \frac{ \BB \times   \JJ}{n e} \right) 
\right\rangle \label{singlecomplete}
\end{align}
\end{widetext}

\paragraph{\bf Neglecting electron mass:} The above equation can be substantially reduced if we completely neglect the effect of electron mass $i.e.$ $\alpha = 0$. In this limit, the exact relation is simply expressed as
\begin{widetext}
\begin{align}
- 4 \varepsilon = &\left\langle \delta \left(\rho \uu \right)  \cdot   \delta \left[ \left( \uu \cdot \nabla \right) \uu \right] + \delta \uu \cdot  \delta \left[ \nabla \cdot \left( \rho \uu \otimes \uu \right) \right] + \gamma \left[ \delta \rho \ \delta \left(\frac{p \theta}{\rho} \right)  + \delta \rho \ \delta \left[ \left( \uu \cdot \nabla \right) e \right] +  \delta \uu  \cdot \ \delta  \left( \rho \  \nabla  e \right) \right] \right. \nonumber \\ 
&\left. - \gamma \left[  \delta \rho_e \ \delta \left(\frac{p_e \chi}{\rho_e} \right) + \delta \rho_e \ \delta \left[ \left( \Jc \cdot \nabla \right) e_e \right] + \delta \left(\Jc\right) \cdot \ \delta \left( \rho_e  \nabla \right) e_e \right]  \right.  \nonumber \\
&\left. + \delta \uu \cdot \delta (\BB \times \JJ) + \delta \jm \cdot \delta \left( \frac{\BB \times \JJ}{\rho} \right) + 2 \delta \JJ \cdot \delta \left[ \BB \times  \left(\uu - \frac{\JJ}{n e} \right) \right] 
\right\rangle \label{single}
\end{align}
\end{widetext}
This exact relation is a generalised version of Hall MHD turbulence where the electron mass is neglected but the electron pressure is not. This is much more realistic model than usual Hall MHD or electron MHD model to probe into the scales comparable or smaller than the electron inertial length. This is the exact relation for the total energy transfer rate in homogeneous and polytropic CHMHD turbulence where electron fluid pressure is not neglected. From this expression, one can easily obtain the exact relations for compressible ordinary MHD turbulence obtained in \citet{B2018} except the thermodynamic energy term which was calculated from an isothermal closure in \citep{B2018} and the self-gravity term which is absent here.  Now, a simplified estimate gives,
\begin{widetext}
\begin{equation}
\frac{\vert \uu \vert }{\vert \JJ/ n e \vert } \approx n e \mu_0 \frac{\vert   \uu \vert }{\vert \nabla \times \BB \vert } \approx \frac{ n e \mu_0  u_{\ell} \ell  }{  \sqrt{\mu_0 n m_i} b_{\ell}} = \left( \frac{\ell}{\lambda_i} \right) \left(\frac{u_{\ell}}{b_{\ell}} \right) {,}
\end{equation}
\end{widetext}
where $\nabla \sim 1/ {\ell}$ with $\ell$ being the length scale which we are interested in, $u_{\ell}$ and $b_{\ell}$ are the velocity and magnetic field fluctuations corresponding to the scale $\ell$.  For typical systems where $u_{\ell} \sim b_{\ell}$, the ratio ${\vert \uu \vert }/{\vert \JJ/ n e \vert } $ is simply given by the factor ${\ell}/{\lambda_i}$ and for ordinary MHD, one is particularly interested in the fluctuation during a very large scale $(\ell >> \lambda_i)$. Hence, at that limit, ${\vert \uu \vert }/{\vert \JJ/ n e \vert } >> 1$ and consequently ${\vert \uu \vert }/{  \vert \alpha \JJ/ n e \vert } >> 1$.  Under the above assumptions, one can finally approximate $\uu_i \approx  \uu_e  \approx \uu $. Interestingly, this does not imply a zero current density $\JJ$ but just indicates a negligible value of the term $\JJ/ ne$ with respect to the fluid velocity $\uu$ at length scales much larger than $\lambda_i$. Due to the same reason, all the terms of Eqn. \eqref{single} containing electron fluid contribution (also containing $\Jc$) can approximately be neglected with respect to similar terms containing single fluid variables (also containing $\uu$). Using the above simplifications, we can easily show that in the limit of CMHD turbulence, the resulting exact relation becomes,
\begin{widetext}
\begin{align}
-4 \varepsilon = \left\langle \delta  \jm   \cdot   \delta \left[ \left( \uu \cdot \nabla \right) \uu \right] + \delta \uu \cdot  \delta \left[ \nabla \cdot \left( \jm  \otimes \uu \right) \right] +  \gamma \delta \rho \ \delta \left[ \frac{p \theta }{\rho} + \left( \uu \cdot \nabla \right) e \right] + \gamma \delta \uu \cdot \delta \left[ \rho \nabla e \right] \right.  \nonumber \\
\left. +  \delta \uu \cdot \delta (\BB \times \JJ) + \delta \jm \cdot \delta \left( \frac{\BB \times \JJ}{\rho} \right) + 2 \delta \JJ \cdot \delta \left( \BB \times  \uu  \right) \right\rangle {.} \label{OMHD}
\end{align} 
\end{widetext}

 
\vspace{1cm}

\subsection{Incompressible single fluid limits}
\vspace{-0.5cm}
In the incompressible single fluid limit, both ion and electron number densities are equal and constant in space and time.  In this limit, we also study two different cases:

\paragraph{\textbf{Without neglecting electron mass}:} As in the previous section, here we also keep the contribution of the electron mass upto first order. The corresponding incompressible single fluid limit can be achieved either by (i) taking the single fluid limit of Eqn.~\eqref{reduction} or (ii) re-writing the equation \eqref{singlecomplete} in the incompressible limit. In the following, we shall only derive the single fluid limit of Eqn.~\eqref{reduction}. For that, we use pre-defined variables:
\begin{widetext}
\begin{equation*}
\uu_i =  \uu + \frac{\alpha}{1 + \alpha} \left( \frac{\JJ}{n e} \right) \approx  \uu + {\alpha} \left( \frac{\JJ}{n e} \right)  \, \, \, \, \text{and}  \, \, \, \, 
\uu_e =  \uu - \frac{1}{1 + \alpha} \left( \frac{\JJ}{n e} \right) \approx  \uu - \left({1 - \alpha}\right) \left( \frac{\JJ}{n e} \right), 
\end{equation*}
\end{widetext}
and hence we obtain
$
\bomega_i \approx  \bomega + {\alpha} \ \boldsymbol{\Omega} \, \, \, \, \text{and}  \, \, \, \, 
\bomega_e \approx  \bomega -  \left({1 - \alpha}\right) \ \boldsymbol{\Omega} , $ where $ \boldsymbol{\Omega} = \nabla \times  \left( \frac{\JJ}{n e} \right)=  \frac{\nabla \times \JJ}{n e} $ (using incompressibility). Now a few steps of straightforward algebra give 
\begin{widetext}
\begin{align*}
 &n m_{i} \delta \uu_i   \cdot   \delta \left( \uu_i \times \bomega_i \right) \approx  \left( 1 - \alpha \right) \rho \ \delta \left( \uu + \alpha  \frac{\JJ}{n e} \right)  \cdot   \delta \left[ \uu \times \bomega  + \alpha {\bf Z} \right]\\
 &n m_{e}  \delta \uu_e   \cdot   \delta \left( \uu_e \times \bomega_e \right) \approx 
 \alpha  \rho \ \delta \left[ \uu - \left(1- \alpha\right)  \frac{\JJ}{n e} \right]  \cdot   \delta \left[ \uu \times \bomega  - \left(1- \alpha \right) {\bf Z}  + \left(1 - 2 \alpha\right) \left(  \frac{\JJ}{n e} \times \boldsymbol{\Omega} \right)  \right], 
\end{align*}
where ${\bf Z}= \left(  \frac{\JJ}{n e} \times \bomega + \uu \times \boldsymbol{\Omega} \right)$. Adding the above two equations, we get
\begin{align}
&\left\langle n m_{i} \delta \uu_i   \cdot   \delta \left( \uu_i \times \bomega_i \right) + n m_{e}  \delta \uu_e   \cdot   \delta \left( \uu_e \times \bomega_e \right) \right\rangle \nonumber  \\ 
&= \rho \left\langle \delta \uu  \cdot   \delta \left( \uu \times \bomega \right) + \alpha \ \delta \left( \uu -  \frac{\JJ}{n e} \right) \cdot \delta \left(  \frac{\JJ}{n e} \times \boldsymbol{\Omega} \right) + \alpha \ \delta \left( \frac{\JJ}{n e} \right) \cdot \delta {\bf Z} \right\rangle .
\end{align}
Again the electromagnetic contribution can be written as (upto first order of $\alpha$)
\begin{align}
 &n e \left\langle  \delta  \uu_i \cdot \delta (\uu_i \times \BB) -   \delta  \uu_e \cdot \delta (\uu_e \times \BB) \right\rangle \nonumber \\
& \approx \left\langle \delta \left( \uu - \frac{\JJ}{n e} \right)  \cdot \delta \left( \JJ \times \BB \right) +  \delta \JJ \cdot \delta \left( \uu \times \BB \right) + 2 \alpha \ \delta \left( \frac{\JJ}{n e} \right) \cdot \delta \left( \JJ \times \BB \right)  \right\rangle
\end{align}
Putting all the contributions together and using Eqn.~\eqref{reduction}, finally the exact relation for incompressible single fluid plasma turbulence is given by 
\begin{align}
2 \varepsilon&= \rho \left\langle \delta \uu  \cdot   \delta \left( \uu \times \bomega \right) + \alpha \ \delta \left( \uu -  \frac{\JJ}{n e} \right) \cdot \delta \left(  \frac{\JJ}{n e} \times \boldsymbol{\Omega} \right) + \alpha \ \delta \left( \frac{\JJ}{n e} \right) \cdot \delta {\bf Z} \right\rangle \nonumber \\
&+ \left\langle \delta \left( \uu - \frac{\JJ}{n e} \right)  \cdot \delta \left( \JJ \times \BB \right) +  \delta \JJ \cdot \delta \left( \uu \times \BB \right) + 2 \alpha  \ \delta \left( \frac{\JJ}{n e} \right) \cdot \delta \left( \JJ \times \BB \right)  \right\rangle \label{incom} 
\end{align}
\end{widetext}
\paragraph{ \textbf{Neglecting electron mass}:} Under the condition where electron mass contribution is entirely ignored, we have $\alpha = 0$. The equation \eqref{incom} is then reduced to 
\begin{widetext}
\begin{align}
2 \varepsilon&= \left\langle \rho \ \delta \uu  \cdot   \delta \left( \uu \times \bomega \right) + \delta \left( \uu - \frac{\JJ}{n e} \right)  \cdot \delta \left( \JJ \times \BB \right) +  \delta \JJ \cdot \delta \left( \uu \times \BB \right)  \right\rangle,  \\
&=\left\langle \rho \ \delta \uu  \cdot   \delta \left( \uu \times \bomega \right) + \delta \uu \cdot \delta \left( \JJ \times \BB \right) +  \delta \JJ \cdot \delta \left[ \left( \uu - \frac{\JJ}{n e} \right)  \times \BB \right]  \right\rangle \label{neglect}
\end{align}
\end{widetext}
which is the exact relation in the incompressible Hall MHD limit. 
For non-relativistic IMHD and IHMHD limits, one can neglect the displacement current term ($\propto \partial {\bf E} / \partial t$) and then write $\nabla \times \BB = \mu_0 \textbf{J} $. Using normalized (to velocity) magnetic field ${\bf b} \equiv \BB / \sqrt{\mu_0 \rho_0} $ and ${\bf j_b} = \nabla \times {\bf b}$ and also assuming incompressibility (n is constant), one can show that ${\JJ}/{n e} = \lambda_i  {\bf j_b}$, where $\lambda_i$ is the ion inertial length scale which is defined as 
\begin{equation}
\lambda_i = \frac{\text{speed of light (c)} }{\text{ion plasma frequency} (\omega_{pi})}= \frac{1}{e}  \sqrt{{\frac{m_i}{n \mu_0}}}
\end{equation}
The equation \eqref{neglect} can therefore be expressed as
\begin{widetext}
\begin{equation}
2 \varepsilon = n M \left\langle   \delta \uu   \cdot   \delta \left( \uu \times \bomega \right)   
+  \left[ \delta  {\bf j_b} \cdot \delta \left\lbrace (\uu - \lambda_i {\bf j_b}) \times \bbb \right\rbrace +  \delta \uu \cdot \delta \left({\bf j_b} \times \bbb \right) \right] \right\rangle , \label{deto}
\end{equation}
\end{widetext}
which is similar to what is obtained (where the constant fluid density is normalized to unity) previously by \citet{B2017} for IHMHD turbulence. For the IMHD limit, we are interested in the length scales much greater than $\lambda_i$, i.e. $\lambda_i \rightarrow 0$ and then Eq.~\eqref{deto} reduces to,
\begin{widetext}
\begin{equation}
2 \varepsilon = \rho  \left\langle   \delta \uu   \cdot   \delta \left( \uu \times \bomega \right)   
+   \delta  {\bf j_b} \cdot \delta  (\uu \times \bbb ) +  \delta \uu \cdot \delta \left({\bf j_b} \times \bbb \right) \right\rangle , \label{deto2}
\end{equation}
\end{widetext}
which is also equal to the exact relation derived by \citep{B2017} for IMHD turbulence, when the uniform density is assumed to be unity.

\subsection{Effect of mean magnetic field}

The total magnetic field $\BB$ at each point can be written as a sum of a mean or uniform background magnetic field $\Bz$ and a fluctuation part. From Eq.~\eqref{incompressible}, one can readily show that the contribution of $\Bz$ to $\varepsilon$ simply vanishes as
$
\left\langle  \rho_{ci0} \ \delta  \uu_i \cdot \delta \uu_i \times \Bz +  \rho_{ce0} \ \delta  \uu_e \cdot \delta \uu_e \times \Bz \right\rangle = 0. 
$
For CTF turbulence, the magnetic contribution due to $\Bz$ is given by 
\begin{widetext}
$$
 \left\langle  \delta (\rho_{ci} \uu_i) \cdot \delta \uu_i \times \Bz + \delta \uu_i \cdot \delta (\rho_{ci} \uu_i) \times \Bz + \delta (\rho_{ce} \uu_e) \cdot \delta \uu_e \times \Bz + \delta \uu_e \cdot \delta (\rho_{ce} \uu_e) \times \Bz  \right\rangle \nonumber= 0 
$$
\end{widetext}
and hence the equation~\eqref{finalexact} remains unaffected by the application of a uniform background field. In the single fluid HMHD limit, the electric field can be expressed in terms of the magnetic field as $\EE = \uu_e \times \BB$ and hence the non-zero contribution comes only from the terms which were originally containing electric field in eqn.~\eqref{finalexact} and the corresponding residual contribution due to $\Bz$ comes to be \\
\begin{widetext}
\begin{align}
& \left\langle - \delta { \JJ} \cdot \delta  \uu_e \times \Bz + \delta \uu \cdot \delta ( n e \uu_e ) \times \Bz - \delta \left(\uu - \frac{\JJ}{n e} \right)  \cdot \delta (n e  \uu_e ) \times \Bz  \right\rangle  \nonumber \\
&= \left\langle - \delta { \JJ} \cdot \delta \left(\uu - \frac{\JJ}{n e} \right) \times \Bz  +  \delta \left(\frac{\JJ}{n e} \right)  \cdot \delta \left( n e  \uu - \JJ \right)  \times \Bz  \right\rangle . \label{52}
\end{align} 
\end{widetext}
In the limit of ordinary MHD, ${\vert \uu \vert }/{\vert \JJ/ n e \vert } >> 1${\modify, and hence} the net contribution due to $\Bz$ comes to be equal to $\left\langle - \delta { \JJ} \cdot \Bz \times \delta \uu   +  \delta \left(\frac{\JJ}{\rho} \right)  \cdot \delta \jm  \times \Bz  \right\rangle$, {\modify which is identical to the contribution of $\Bz$ to the turbulent energy flux rate in compressible MHD turbulence as obtained previously \citep{B2018}.}
\vspace{-0.7cm}
\section{Conclusions}\label{conclusions}
\vspace{-0.5cm}
For fully developed turbulence, we derive the exact relation for a 3D compressible two-fluid plasma model. Equation \eqref{finalexact} can be used to compute the total energy dissipation or transfer rate in a weakly ionized plasma which can mostly be seen in the dilute astrophysical plasmas like cold molecular clouds or protoplanetary disks. Till now, astrophysical turbulence has been mostly studied using hydrodynamic simulation and very few times using MHD single fluid model (for a detailed review, see \citep{Schmidt}) which is a simplistic assumption for an astrophysical medium. This current work will facilitate the study of the astrophysical plasma turbulence to a large extent. In particular, two-fluid compressible plasma turbulence model can be expected to give a reliable estimate of the star formation efficiency in the cold molecular clouds which can then be studied using appropriate numerical simulation. In addition, owing to this law, the effect of background magnetic field in the turbulent star forming regions can also be studied using CTF plasma turbulence simulation. An interesting study will be to understand at which scale the electric field contribution gets converted to the background magnetic field contribution as obtained in the Eqn.~\eqref{52}. Furthermore, this exact relation will help estimate the energy dissipation or heating rate of the ionosphere due to ionospheric plasma turbulence. Note that, in partially or weakly ionized plasmas, the neutral particles constitute an essential parts of the flow dynamics. However, their contribution can be easily included in the kinetic and the thermodynamic energy contribution. For solar wind, a CTF model of plasma includes the effect of ion and electron inertia in the scaling of turbulence, and thus generalizes previous results obtained for incompressible and compressible MHD and HMHD \citep{P1998b, Ga2008, B2013, B2014, A2017b}. {\modify Another interesting fact is that our exact relation does not a priori assume any isotropy. It is known \citep{Miura, Comisso16} that turbulence eddies in the form of current sheets and vorticity layers can become more and more anisotropic and finally unstable to plasma instabilities when a large inertial range is considered. Although these instabilities can steepen the magnetic and kinetic energy spectra \citep{Comisso18}, the form of the exact relation is expected to be unaltered provided the $\varepsilon$ will not be a scale invariant quantity any longer. }

 For scales larger than dissipative scales, Eq.~\eqref{finalexact} implies a specific scaling for structure functions of the velocity of each species and of the magnetic field in a non-trivial way, which can be related to the energy dissipation rate $\varepsilon$ at the smallest scales. Therefore, it provides a way to test whether a range of scales in a plasma is inertial or dissipative. Over the last years, the sustained increase in the spatial and temporal resolution of space missions such as Cluster (ESA) or the new NASA MMS (Magnetospheric MultiScale) mission has opened the possibility to investigate small-scale plasma phenomena as never before. The exact laws derived here allow investigation of the nature of turbulent magnetic field fluctuations at a broad range of scales in space plasmas, and will be essential to understand the nonlinear nature of turbulence at the electron scales in the solar wind.

\section{Acknowledgements}
N.A.  thanks Daniel O. G\'omez for useful discussions. S.B. acknowledges research grant from DST INSPIRE fellowship (DST/PHY/2017514). N.A. acknowledges financial support from CNRS/CONICET Laboratoire International Associ\'e (LIA) MAGNETO.


\begin{thebibliography}{22}%
	\makeatletter
	\providecommand \@ifxundefined [1]{%
		\@ifx{#1\undefined}
	}%
	\providecommand \@ifnum [1]{%
		\ifnum #1\expandafter \@firstoftwo
		\else \expandafter \@secondoftwo
		\fi
	}%
	\providecommand \@ifx [1]{%
		\ifx #1\expandafter \@firstoftwo
		\else \expandafter \@secondoftwo
		\fi
	}%
	\providecommand \natexlab [1]{#1}%
	\providecommand \enquote  [1]{``#1''}%
	\providecommand \bibnamefont  [1]{#1}%
	\providecommand \bibfnamefont [1]{#1}%
	\providecommand \citenamefont [1]{#1}%
	\providecommand \href@noop [0]{\@secondoftwo}%
	\providecommand \href [0]{\begingroup \@sanitize@url \@href}%
	\providecommand \@href[1]{\@@startlink{#1}\@@href}%
	\providecommand \@@href[1]{\endgroup#1\@@endlink}%
	\providecommand \@sanitize@url [0]{\catcode `\\12\catcode `\$12\catcode
		`\&12\catcode `\#12\catcode `\^12\catcode `\_12\catcode `\%12\relax}%
	\providecommand \@@startlink[1]{}%
	\providecommand \@@endlink[0]{}%
	\providecommand \url  [0]{\begingroup\@sanitize@url \@url }%
	\providecommand \@url [1]{\endgroup\@href {#1}{\urlprefix }}%
	\providecommand \urlprefix  [0]{URL }%
	\providecommand \Eprint [0]{\href }%
	\providecommand \doibase [0]{http://dx.doi.org/}%
	\providecommand \selectlanguage [0]{\@gobble}%
	\providecommand \bibinfo  [0]{\@secondoftwo}%
	\providecommand \bibfield  [0]{\@secondoftwo}%
	\providecommand \translation [1]{[#1]}%
	\providecommand \BibitemOpen [0]{}%
	\providecommand \bibitemStop [0]{}%
	\providecommand \bibitemNoStop [0]{.\EOS\space}%
	\providecommand \EOS [0]{\spacefactor3000\relax}%
	\providecommand \BibitemShut  [1]{\csname bibitem#1\endcsname}%
	\let\auto@bib@innerbib\@empty
	\bibitem [{\citenamefont {Bruno}\ and\ \citenamefont
		{Carbone}(2005)}]{Bruno05}%
	\BibitemOpen
	\bibfield  {author} {\bibinfo {author} {\bibfnamefont {R.}\ \bibnamefont
			{Bruno}}\ and\ \bibinfo {author} {\bibfnamefont {V.}\ \bibnamefont
			{Carbone}},\ }\href@noop {} \bibfield  {title} {\enquote {\bibinfo {title} {The Solar Wind as a Turbulence Laboratory},}\ }\href@noop {} {\bibfield  {journal} {\bibinfo  {journal} {Living Reviews in Solar Physics}\ }\textbf {\bibinfo {volume} {2}},\ \bibinfo {pages} {4}(\bibinfo {year} {2005})}\BibitemShut {NoStop}%
	\bibitem [{\citenamefont {Matthaeus}\ and\ \citenamefont
		{Goldstein}(1982)}]{M1982}%
	\BibitemOpen
	\bibfield  {author} {\bibinfo {author} {\bibfnamefont {W.~H.}\ \bibnamefont
			{Matthaeus}}\ and\ \bibinfo {author} {\bibfnamefont {M.~L.}\ \bibnamefont
			{Goldstein}},\ }\href@noop {} \bibfield  {title} {\enquote {\bibinfo {title} {Measurement of the rugged invariants of magnetohydrodynamic turbulence in the solar wind},}\ }\href@noop {} {\bibfield  {journal} {\bibinfo  {journal} {Journal of
				Geophysical Research}\ }\textbf {\bibinfo {volume} {87}},\ \bibinfo {pages} {6011}
		(\bibinfo {year} {1982})}\BibitemShut {NoStop}%
	\bibitem [{\citenamefont {Frisch}(1995)}]{F1995}%
	\BibitemOpen
	\bibfield  {author} {\bibinfo {author} {\bibfnamefont {U.}~\bibnamefont
			{Frisch}},\ }\href@noop {} {\emph {\bibinfo {title} {Turbulence: The Legacy
				of A. N. Kolmogorov}}}\ (\bibinfo  {publisher} {Cambridge University
		Press.},\ \bibinfo {year} {1995})\BibitemShut {NoStop}%
		\bibitem [{\citenamefont {Politano}\ and\ \citenamefont
		{Pouquet}(1998{\natexlab{a}})}]{P1998a}%
	\BibitemOpen
	\bibfield  {author} {\bibinfo {author} {\bibfnamefont {H.}~\bibnamefont
			{Politano}}\ and\ \bibinfo {author} {\bibfnamefont {A.}~\bibnamefont
			{Pouquet}},\ }\bibfield  {title} {\enquote {\bibinfo {title} {von
				k{\'a}rm{\'a}n--howarth equation for magnetohydrodynamics and its
				consequences on third-order longitudinal structure and correlation
				functions},}\ }\href@noop {} {\bibfield  {journal} {\bibinfo  {journal}
			{Physical Review E}\ }\textbf {\bibinfo {volume} {57}},\ \bibinfo {pages}
		{R21} (\bibinfo {year} {1998}{\natexlab{a}})}\BibitemShut {NoStop}%
	\bibitem [{\citenamefont {von K\'arm\'an}\ and\ \citenamefont
		{Howarth}(1938)}]{vkh1938}%
	\BibitemOpen
	\bibfield  {author} {\bibinfo {author} {\bibfnamefont {T.}\
			\bibnamefont {von K\'arm\'an}}\ and\ \bibinfo {author} {\bibfnamefont
			{L.}~\ \bibnamefont {Howarth}},\ }\bibfield  {title} {\enquote {\bibinfo
			{title} {On the statistical theory of isotropic turbulence},}\ }\href
	{\doibase 10.1098/rspa.1938.0013} {\bibfield  {journal} {\bibinfo  {journal}
			{Proceedings of the Royal Society of London A: Mathematical, Physical and
				Engineering Sciences}\ }\textbf {\bibinfo {volume} {164}},\ \bibinfo {pages}
		{192--215} (\bibinfo {year} {1938})}\BibitemShut {NoStop}%
	\bibitem [{\citenamefont {Kolmogorov}(1941a)}]{K1941a}%
	\BibitemOpen
	\bibfield  {author} {\bibinfo {author} {\bibfnamefont {A.~N.}\ \bibnamefont
			{Kolmogorov}},\ }\href@noop {} \bibfield  {title} {\enquote {\bibinfo {title} {The Local Structure of Turbulence in Incompressible Viscous Fluid for Very Large Reynolds' Numbers},}\ }\href@noop {} {\bibfield  {journal} {\bibinfo  {journal}
			{Akademiia Nauk SSSR Doklady}\ }\textbf {\bibinfo {volume} {30}},\ \bibinfo
		{pages} {301} (\bibinfo {year} {1941a})}\BibitemShut {NoStop}%
	\bibitem [{\citenamefont {Kolmogorov}(1941b)}]{K1941b}%
	\BibitemOpen
	\bibfield  {author} {\bibinfo {author} {\bibfnamefont {A.~N.}\ \bibnamefont
			{Kolmogorov}},\ }\href@noop {}  \bibfield  {title} {\enquote {\bibinfo {title} {Energy dissipation in locally isotropic turbulence.},}\ }\href@noop {} {\bibfield  {journal} {\bibinfo  {journal}
			{Comptes Rendus de l'Acad\'emie des Sciences de l'URSS}\ }\textbf {\bibinfo {volume} {32}},\ \bibinfo {pages} {19}
		(\bibinfo {year} {1941b})}\BibitemShut {NoStop}%
		\bibitem [{\citenamefont {Monin}\ and\ \citenamefont {Yaglom}(1975)}]{MY1975}%
		\BibitemOpen
	\bibfield  {author} {\bibinfo {author} {\bibfnamefont {A.~S.}\ \bibnamefont
			{Monin}}\ and\ \bibinfo {author} {\bibfnamefont {A.~M.}\ \bibnamefont
			{Yaglom}},\ }\href@noop {} {\emph {\bibinfo {title} {Statistical Fluid
				Mechanics: Mechanics of Turbulence}}},\ Vol.~\bibinfo {volume} {2}\ (\bibinfo
	{publisher} {Cambridge, MA: MIT Press.},\ \bibinfo {year}
	{1975})\BibitemShut {NoStop}%
	\bibitem [{\citenamefont {Banerjee}\ \emph {et~al.}(2016)\citenamefont
		{Banerjee}, \citenamefont {Hadid}, \citenamefont {Sahraoui},\ and\
		\citenamefont {Galtier}}]{B2016c}%
	\BibitemOpen
	\bibfield  {author} {\bibinfo {author} {\bibfnamefont {S.}\
			\bibnamefont {Banerjee}}, \bibinfo {author} {\bibfnamefont {L.~Z.}\
			\bibnamefont {Hadid}}, \bibinfo {author} {\bibfnamefont {F.}\ \bibnamefont
			{Sahraoui}}, \ and\ \bibinfo {author} {\bibfnamefont {S.}\
			\bibnamefont {Galtier}},\ }\bibfield  {title} {\enquote {\bibinfo {title}
			{Scaling of compressible magnetohydrodynamic turbulence in the fast solar
				wind},}\ }\href@noop {} {\bibfield  {journal} {\bibinfo  {journal} {The
				Astrophysical Journal Letters}\ }\textbf {\bibinfo {volume} {829}},\ \bibinfo
		{pages} {L27} (\bibinfo {year} {2016})}\BibitemShut {NoStop}%
	\bibitem [{\citenamefont {Politano}\ and\ \citenamefont
		{Pouquet}(1998{\natexlab{b}})}]{P1998b}%
	\BibitemOpen
	\bibfield  {author} {\bibinfo {author} {\bibfnamefont {H.}\
			\bibnamefont {Politano}}\ and\ \bibinfo {author} {\bibfnamefont {A.}\
			\bibnamefont {Pouquet}},\ }\bibfield  {title} {\enquote {\bibinfo {title}
			{Dynamical length scales for turbulent magnetized flows},}\ }\href@noop {}
	{\bibfield  {journal} {\bibinfo  {journal} {Geophysical Research Letters}\
		}\textbf {\bibinfo {volume} {25}},\ \bibinfo {pages} {273--276} (\bibinfo
		{year} {1998}{\natexlab{b}})}\BibitemShut {NoStop}%
	\bibitem [{\citenamefont {Politano}\ \emph {et~al.}(2003)\citenamefont
		{Politano}, \citenamefont {Gomez},\ and\ \citenamefont {Pouquet}}]{P2003}%
	\BibitemOpen
	\bibfield  {author} {\bibinfo {author} {\bibfnamefont {H.}~\bibnamefont
			{Politano}}, \bibinfo {author} {\bibfnamefont {T.}~\bibnamefont {Gomez}}, \
		and\ \bibinfo {author} {\bibfnamefont {A.}~\bibnamefont {Pouquet}},\
	}\href@noop {} \bibfield  {title} {\enquote {\bibinfo {title}
			{von K\`arm\`an–Howarth relationship for helical magnetohydrodynamic flows},}\ }\href@noop {}{\bibfield  {journal} {\bibinfo  {journal} {Physical Review E}\
		}\textbf {\bibinfo {volume} {68}},\ \bibinfo {pages} {026315} (\bibinfo
		{year} {2003})}\BibitemShut {NoStop}%
	\bibitem [{\citenamefont {Galtier}(2008)}]{Ga2008e}%
	\BibitemOpen
	\bibfield  {author} {\bibinfo {author} {\bibfnamefont {S.}\
			\bibnamefont {Galtier}},\ }\bibfield  {title} {\enquote {\bibinfo {title}
			{Exact scaling laws for 3D electron MHD turbulence},}\
	}\href@noop {} {\bibfield  {journal} {\bibinfo  {journal} {Journal of Geophysical Research: Space Physics}\
		}\textbf {\bibinfo {volume} {113}},\ \bibinfo {pages} {A1} (\bibinfo
		{year} {2008})}\BibitemShut {NoStop}%
	\bibitem [{\citenamefont {Galtier}(2008)}]{Ga2008}%
	\BibitemOpen
	\bibfield  {author} {\bibinfo {author} {\bibfnamefont {S.}\
			\bibnamefont {Galtier}},\ }\bibfield  {title} {\enquote {\bibinfo {title}
			{von k{\'a}rm{\'a}n--howarth equations for hall magnetohydrodynamic flows},}\
	}\href@noop {} {\bibfield  {journal} {\bibinfo  {journal} {Physical Review E}\
		}\textbf {\bibinfo {volume} {77}},\ \bibinfo {pages} {015302} (\bibinfo
		{year} {2008})}\BibitemShut {NoStop}%
\bibitem [{\citenamefont {Meyrand}(2010)}]{Meyrand10}%
	\BibitemOpen
	\bibfield  {author} {\bibinfo {author} {\bibfnamefont {R.}\
			\bibnamefont {Meyrand}} and\ \bibinfo {author} {\bibfnamefont {S.}\ \bibnamefont
			{Galtier}},\ }\bibfield  {title} {\enquote {\bibinfo {title}
			{A universal law for solar-wind turbulence at electron scales},}\
	}\href@noop {} {\bibfield  {journal} {\bibinfo  {journal} {The Astrop[hysical Journal}\
		}\textbf {\bibinfo {volume} {721}},\ \bibinfo {pages} {1421-1424} (\bibinfo
		{year} {2010})}\BibitemShut {NoStop}%
\bibitem [{\citenamefont {Banerjee}\ and\ \citenamefont
		{Galtier}(2017)}]{B2017}%
	\BibitemOpen
	\bibfield  {author} {\bibinfo {author} {\bibfnamefont {S.}\
			\bibnamefont {Banerjee}}\ and\ \bibinfo {author} {\bibfnamefont
			{S.}\ \bibnamefont {Galtier}},\ }\bibfield  {title} {\enquote
		{\bibinfo {title} {An alternative formulation for exact scaling relations in hydrodynamic and magnetohydrodynamic turbulence},}\ }\href@noop {} {\bibfield
		{journal} {\bibinfo  {journal} {Journal of Physics A: Mathematical and
				Theoretical}\ }\textbf {\bibinfo {volume} {50}},\ \bibinfo {pages} {015501}
		(\bibinfo {year} {2017})}\BibitemShut {NoStop}%
		\bibitem [{\citenamefont {Andres}\ and\ \citenamefont
		{Banerjee}(2019)}]{AndresBan}%
	\BibitemOpen
	\bibfield  {author} {\bibinfo {author} {\bibfnamefont {N.}\
			\bibnamefont {Andres}}\ and\ \bibinfo {author} {\bibfnamefont
			{S.}\ \bibnamefont {Banerjee}},\ }\bibfield  {title} {\enquote
		{\bibinfo {title} {Statistics of incompressible hydrodynamic turbulence: An alternative approach},}\ }\href@noop {} {\bibfield
		{journal} {\bibinfo  {journal} {Physical Review Fluids}\ }\textbf {\bibinfo {volume} {4}},\ \bibinfo {pages} {024603}
		(\bibinfo {year} {2019})}\BibitemShut {NoStop}%
\bibitem [{\citenamefont {Ferrand}\ \emph {et~al.}(2019)\citenamefont {Ferrand},
		\citenamefont {Galtier}, \citenamefont {Sahraoui}, \citenamefont {Meyrand}, \citenamefont {Andres},\ and\ \citenamefont
		{Banerjee}}]{Ferrand2019}%
	\BibitemOpen
	\bibfield  {author} {\bibinfo {author} {\bibfnamefont {R.}\ \bibnamefont
			{Ferrand}}, \bibinfo {author} {\bibfnamefont {S.}\ \bibnamefont {Galtier}},
		\bibinfo {author} {\bibfnamefont {F.}\ \bibnamefont {Sahraoui}}, \bibinfo {author} {\bibfnamefont {R.}\ \bibnamefont {Meyrand}}, \bibinfo {author} {\bibfnamefont {N.}\ \bibnamefont {Andres}} \ and\
		\bibinfo {author} {\bibfnamefont {S.}\ \bibnamefont {Banerjee}},\
	}\bibfield  {title} {\enquote {\bibinfo {title} {On exact laws in incompressible Hall MHD turbulence},}\ }\href@noop {}
	{\bibfield  {journal} {\bibinfo  {journal} {The Astrophysical Journal}\ }\textbf
		{\bibinfo {volume} {881}},\ \bibinfo {pages} {50} (\bibinfo {year}
		{2019})}\BibitemShut {NoStop}%
	\bibitem [{\citenamefont {Galtier}\ and\ \citenamefont
		{Banerjee}(2011)}]{Ga2011}%
	\BibitemOpen
	\bibfield  {author} {\bibinfo {author} {\bibfnamefont {S.}\
			\bibnamefont {Galtier}}\ and\ \bibinfo {author} {\bibfnamefont {S.}\
			\bibnamefont {Banerjee}},\ }\bibfield  {title} {\enquote {\bibinfo {title}
			{Exact relation for correlation functions in compressible isothermal
				turbulence},}\ }\href@noop {} {\bibfield  {journal} {\bibinfo  {journal}
			{Physical Review Letters}\ }\textbf {\bibinfo {volume} {107}},\ \bibinfo
		{pages} {134501} (\bibinfo {year} {2011})}\BibitemShut {NoStop}%
	\bibitem [{\citenamefont {Banerjee}\ and\ \citenamefont
		{Galtier}(2014)}]{B2014}%
	\BibitemOpen
	\bibfield  {author} {\bibinfo {author} {\bibfnamefont {S.}\
			\bibnamefont {Banerjee}}\ and\ \bibinfo {author} {\bibfnamefont
			{S.}\ \bibnamefont {Galtier}},\ }\bibfield  {title} {\enquote
		{\bibinfo {title} {A kolmogorov-like exact relation for compressible
				polytropic turbulence},}\ }\href@noop {} {\bibfield  {journal} {\bibinfo
			{journal} {Journal of Fluid Mechanics}\ }\textbf {\bibinfo {volume} {742}},\
		\bibinfo {pages} {230--242} (\bibinfo {year} {2014})}\BibitemShut {NoStop}%
	\bibitem [{\citenamefont {Banerjee}\ and\ \citenamefont
		{Galtier}(2013)}]{B2013}%
	\BibitemOpen
	\bibfield  {author} {\bibinfo {author} {\bibfnamefont {S.}\
			\bibnamefont {Banerjee}}\ and\ \bibinfo {author} {\bibfnamefont
			{S.}\ \bibnamefont {Galtier}},\ }\bibfield  {title} {\enquote
		{\bibinfo {title} {Exact relation with two-point correlation functions and
				phenomenological approach for compressible magnetohydrodynamic turbulence},}\
	}\href@noop {} {\bibfield  {journal} {\bibinfo  {journal} {Physical Review
				E}\ }\textbf {\bibinfo {volume} {87}},\ \bibinfo {pages} {013019} (\bibinfo
		{year} {2013})}\BibitemShut {NoStop}%
	\bibitem [{\citenamefont {Andr{\'e}s}\ and\ \citenamefont
		{Sahraoui}(2017)}]{A2017b}%
	\BibitemOpen
	\bibfield  {author} {\bibinfo {author} {\bibfnamefont {N.}\ \bibnamefont
			{Andr{\'e}s}}\ and\ \bibinfo {author} {\bibfnamefont {F.}\ \bibnamefont
			{Sahraoui}},\ }\bibfield  {title} {\enquote {\bibinfo {title} {Alternative
				derivation of exact law for compressible and isothermal magnetohydrodynamics
				turbulence},}\ }\href@noop {} {\bibfield  {journal} {\bibinfo  {journal}
			{Physical Review E}\ }\textbf {\bibinfo {volume} {96}},\ \bibinfo {pages}
		{053205} (\bibinfo {year} {2017})}\BibitemShut {NoStop}%
		\bibitem [{\citenamefont {Andrés}\ \emph {et~al.}(2018)\citenamefont
		{Andrés}, \citenamefont {Galtier}, \and \ \citenamefont {Sahraoui}}]{A2018}%
	\BibitemOpen
	\bibfield  {author} {\bibinfo {author} {\bibfnamefont {N.}~\bibnamefont
			{Andrés}}, \bibinfo {author} {\bibfnamefont {S.}~\bibnamefont {Galtier}},\ and\
		\bibinfo {author} {\bibfnamefont {F.}~\bibnamefont {Sahraoui}},\ }\bibfield  {title}
	{\enquote {\bibinfo {title} {Exact law for homogeneous compressible Hall magnetohydrodynamics turbulence},}\ }\href {\doibase 10.1103/PhysRevE.97.013204
	} {\bibfield  {journal} {\bibinfo  {journal}
			{Physical Review E}\ }\textbf {\bibinfo {volume} {97}},\ \bibinfo
		{pages} {013204} (\bibinfo {year} {2018})}\BibitemShut {NoStop}%
			\bibitem [{\citenamefont {Banerjee}\ and\ \citenamefont
		{Kritsuk}(2018{\natexlab{a}})}]{B2017k}%
	\BibitemOpen
	\bibfield  {author} {\bibinfo {author} {\bibfnamefont {S.}\
			\bibnamefont {Banerjee}}\ and\ \bibinfo {author} {\bibfnamefont {A.~G.}\
			\bibnamefont {Kritsuk}},\ }\bibfield  {title} {\enquote {\bibinfo {title}
			{Exact relations for energy transfer in self-gravitating isothermal turbulence},}\ }\href {\doibase
		10.1103/PhysRevE.96.053116} {\bibfield  {journal} {\bibinfo  {journal} {Physical
				Review E}\ }\textbf {\bibinfo {volume} {96}},\ \bibinfo {pages} {053116}
		(\bibinfo {year} {2018}{\natexlab{a}})}\BibitemShut {NoStop}%
		\bibitem [{\citenamefont {Aluie}(2011)}]{A2011}%
	\BibitemOpen
	\bibfield  {author} {\bibinfo {author} {\bibfnamefont {H.}\ \bibnamefont
			{Aluie}},\ }\bibfield  {title} {\enquote {\bibinfo {title} {Compressible
				turbulence: the cascade and its locality},}\ }\href@noop {} {\bibfield
		{journal} {\bibinfo  {journal} {Physical Review Letters}\ }\textbf {\bibinfo
			{volume} {106}},\ \bibinfo {pages} {174502} (\bibinfo {year}
		{2011})}\BibitemShut {NoStop}%
		\bibitem [{\citenamefont {Banerjee}\ and\ \citenamefont
		{Kritsuk}(2018{\natexlab{a}})}]{B2018}%
	\BibitemOpen
	\bibfield  {author} {\bibinfo {author} {\bibfnamefont {S.}\
			\bibnamefont {Banerjee}}\ and\ \bibinfo {author} {\bibfnamefont {A.~G.}\
			\bibnamefont {Kritsuk}},\ }\bibfield  {title} {\enquote {\bibinfo {title}
			{Energy transfer in compressible magnetohydrodynamic turbulence for
				isothermal self-gravitating fluids},}\ }\href {\doibase
		10.1103/PhysRevE.97.023107} {\bibfield  {journal} {\bibinfo  {journal} {Physical
				Review E}\ }\textbf {\bibinfo {volume} {97}},\ \bibinfo {pages} {023107}
		(\bibinfo {year} {2018}{\natexlab{a}})}\BibitemShut {NoStop}%
		\bibitem [{\citenamefont {Hadid}\ \emph {et~al.}(2017)\citenamefont {Hadid},
		\citenamefont {Sahraoui},\ and\ \citenamefont {Galtier}}]{H2017a}%
	\BibitemOpen
	\bibfield  {author} {\bibinfo {author} {\bibfnamefont {L.~Z.}\bibnamefont
			{Hadid}}, \bibinfo {author} {\bibfnamefont {F}~\bibnamefont {Sahraoui}}, \
		and\ \bibinfo {author} {\bibfnamefont {S}~\bibnamefont {Galtier}},\
	}\bibfield  {title} {\enquote {\bibinfo {title} {Energy cascade rate in
				compressible fast and slow solar wind turbulence},}\ }\href@noop {}
	{\bibfield  {journal} {\bibinfo  {journal} {The Astrophysical Journal}\
		}\textbf {\bibinfo {volume} {838}},\ \bibinfo {pages} {9} (\bibinfo {year}
		{2017})}\BibitemShut {NoStop}%
	\bibitem [{\citenamefont {Hadid}\ \emph {et~al.}(2018)\citenamefont {Hadid},
		\citenamefont {Sahraoui}, \citenamefont {Galtier},\ and\ \citenamefont
		{Huang}}]{H2017b}%
	\BibitemOpen
	\bibfield  {author} {\bibinfo {author} {\bibfnamefont {L.~Z.}\ \bibnamefont
			{Hadid}}, \bibinfo {author} {\bibfnamefont {F.}\ \bibnamefont {Sahraoui}},
		\bibinfo {author} {\bibfnamefont {S.}\ \bibnamefont {Galtier}}, \ and\
		\bibinfo {author} {\bibfnamefont {S.}\ \bibnamefont {Huang}},\
	}\bibfield  {title} {\enquote {\bibinfo {title} {Compressible
				magnetohydrodynamic turbulence in the earth's magnetosheath: estimation of the energy cascade rate using in situ spacecraft data},}\ }\href@noop {}
	{\bibfield  {journal} {\bibinfo  {journal} {Physical Review Letters}\ }\textbf
		{\bibinfo {volume} {120}},\ \bibinfo {pages} {055102} (\bibinfo {year}
		{2018})}\BibitemShut {NoStop}%
	\bibitem [{\citenamefont {Andrés}\ \emph {et~al.}(2019)\citenamefont
		{Andrés}, \citenamefont {Sahraoui}, \citenamefont {Galtier}, \citenamefont {Hadid}, \citenamefont {Ferrand}, \and \ \citenamefont {Huang}}]{A2019PRL}%
	\BibitemOpen
	\bibfield  {author} {\bibinfo {author} {\bibfnamefont {N.}~\bibnamefont
			{Andrés}}, \bibinfo {author} {\bibfnamefont {F.}~\bibnamefont {Sahraoui}}, \bibinfo {author} {\bibfnamefont {S.}~\bibnamefont {Galtier}}, \bibinfo {author} {\bibfnamefont {L. Z.}~\bibnamefont {Hadid}}, \bibinfo {author} {\bibfnamefont {R.}~\bibnamefont {Ferrand}},\ and\
		\bibinfo {author} {\bibfnamefont {S. Y.}~\bibnamefont {Huang}},\ }\bibfield  {title}
	{\enquote {\bibinfo {title} {Energy Cascade Rate Measured in a Collisionless Space Plasma with MMS Data and Compressible Hall Magnetohydrodynamic Turbulence Theory},}\ }\href {\doibase 10.1103/PhysRevLett.123.245101
	} {\bibfield  {journal} {\bibinfo  {journal}
			{Physical Review Letters}\ }\textbf {\bibinfo {volume} {123}},\ \bibinfo
		{pages} {245101} (\bibinfo {year} {2019})}\BibitemShut {NoStop}%
	\bibitem [{\citenamefont {Kritsuk}\ \emph {et~al.}(2013)\citenamefont
		{Kritsuk}, \citenamefont {Wagner},\ and\ \citenamefont {Norman}}]{K2013}%
	\BibitemOpen
	\bibfield  {author} {\bibinfo {author} {\bibfnamefont {A.~G.}\ \bibnamefont
			{Kritsuk}}, \bibinfo {author} {\bibfnamefont {R.}~\bibnamefont {Wagner}}, \
		and\ \bibinfo {author} {\bibfnamefont {M.~L.}\ \bibnamefont {Norman}},\
	}\href@noop {}\bibfield  {title} {\enquote {\bibinfo {title} {Energy cascade and scaling in supersonic isothermal turbulence},}\ }\href@noop {} {\bibfield  {journal} {\bibinfo  {journal} {Journal of Fluid
				Mechanics}\ }\textbf {\bibinfo {volume} {729}},\ \bibinfo {pages} {R1}
		(\bibinfo {year} {2013})}\BibitemShut {NoStop}%
	\bibitem [{\citenamefont {Andrés}\ \emph {et~al.}(2018)\citenamefont
		{Andrés}, \citenamefont {Sahraoui}, \citenamefont {Galtier}, \citenamefont
		{Hadid}, \citenamefont {Dmitruk},\ and\ \citenamefont {Mininni}}]{A2018b}%
	\BibitemOpen
	\bibfield  {author} {\bibinfo {author} {\bibfnamefont {N.}~\bibnamefont
			{Andrés}}, \bibinfo {author} {\bibfnamefont {F.}~\bibnamefont {Sahraoui}},
		\bibinfo {author} {\bibfnamefont {S.}~\bibnamefont {Galtier}}, \bibinfo
		{author} {\bibfnamefont {L.~Z.}\ \bibnamefont {Hadid}}, \bibinfo {author}
		{\bibfnamefont {P.}~\bibnamefont {Dmitruk}}, \ and\ \bibinfo {author}
		{\bibfnamefont {P.~D.}\ \bibnamefont {Mininni}},\ }\bibfield  {title}
	{\enquote {\bibinfo {title} {Energy cascade rate in isothermal compressible
				magnetohydrodynamic turbulence},}\ }\href {\doibase
		10.1017/S0022377818000788} {\bibfield  {journal} {\bibinfo  {journal}
			{Journal of Plasma Physics}\ }\textbf {\bibinfo {volume} {84}},\ \bibinfo
		{pages} {905840404} (\bibinfo {year} {2018})}\BibitemShut {NoStop}%
\bibitem [{\citenamefont {Andr{\'e}s}\ \emph
		{et~al.}(2016{\natexlab{a}})\citenamefont {Andr{\'e}s}, \citenamefont
		{Mininni}, \citenamefont {Dmitruk},\ and\ \citenamefont {Gomez}}]{A2016b}%
	\BibitemOpen
	\bibfield  {author} {\bibinfo {author} {\bibfnamefont {N.}\ \bibnamefont
			{Andr{\'e}s}}, \bibinfo {author} {\bibfnamefont {P.~D.}\ \bibnamefont
			{Mininni}}, \bibinfo {author} {\bibfnamefont {P.}\ \bibnamefont
			{Dmitruk}}, \ and\ \bibinfo {author} {\bibfnamefont {D.~O.}\
			\bibnamefont {Gomez}},\ }\bibfield  {title} {\enquote {\bibinfo {title} {von
				K{\'a}rm{\'a}n--Howarth equation for three-dimensional two-fluid plasmas},}\
	}\href@noop {} {\bibfield  {journal} {\bibinfo  {journal} {Physical Review
				E}\ }\textbf {\bibinfo {volume} {93}},\ \bibinfo {pages} {063202} (\bibinfo
		{year} {2016}{\natexlab{a}})}\BibitemShut {NoStop}%
		\bibitem [{\citenamefont {Andr{\'e}s}\ \emph
		{et~al.}(2016{\natexlab{b}})\citenamefont {Andr{\'e}s}, \citenamefont
		{Galtier},\ and\ \citenamefont {Sahraoui}}]{A2016c}%
	\BibitemOpen
	\bibfield  {author} {\bibinfo {author} {\bibfnamefont {N.}\ \bibnamefont
			{Andr{\'e}s}}, \bibinfo {author} {\bibfnamefont {S.}\ \bibnamefont
			{Galtier}}, \ and\ \bibinfo {author} {\bibfnamefont {F.}\ \bibnamefont
			{Sahraoui}},\ }\bibfield  {title} {\enquote {\bibinfo {title} {Exact scaling
				laws for helical three-dimensional two-fluid turbulent plasmas},}\
	}\href@noop {} {\bibfield  {journal} {\bibinfo  {journal} {Physical Review
				E}\ }\textbf {\bibinfo {volume} {94}},\ \bibinfo {pages} {063206} (\bibinfo
		{year} {2016}{\natexlab{b}})}\BibitemShut {NoStop}%
		\bibitem [{\citenamefont {Bittencourt}(2004)}]{Bittencourt}%
	\BibitemOpen
	\bibfield  {author} {\bibinfo {author} {\bibfnamefont {J. A.}~\bibnamefont
			{Bittencourt}},\ }\href@noop {} {\emph {\bibinfo {title} {Fundamentals of Plasma Physics}}}\ (\bibinfo  {publisher} {Springer},\ \bibinfo {year} {2004})\BibitemShut {NoStop}%
\bibitem [{\citenamefont {Zweibel}\ and\ \citenamefont {McKee}(1995)}]{Z1995}%
	\BibitemOpen
	\bibfield  {author} {\bibinfo {author} {\bibfnamefont {E.~G.}\ \bibnamefont
			{Zweibel}}\ and\ \bibinfo {author} {\bibfnamefont {C.~F.}\
			\bibnamefont {McKee}},\ }\bibfield  {title} {\enquote {\bibinfo {title}
			{Equiparatition of energy for turbulent astrophysical fluids: Accounting for
				the unseen energy in molecular clouds},}\ }\href@noop {} {\bibfield
		{journal} {\bibinfo  {journal} {The Astrophysical Journal}\ }\textbf
		{\bibinfo {volume} {439}},\ \bibinfo {pages} {779--792} (\bibinfo {year}
		{1995})}\BibitemShut {NoStop}%
	\bibitem [{\citenamefont {Landau}\ and\ \citenamefont
		{Lifshitz}(2013)}]{Lan2013}%
	\BibitemOpen
	\bibfield  {author} {\bibinfo {author} {\bibfnamefont {L. D.}\
			\bibnamefont {Landau}}\ and\ \bibinfo {author} {\bibfnamefont
			{E. M.}\ \bibnamefont {Lifshitz}},\ }\bibfield  {title} {\enquote
		{\bibinfo {title} {Fluid Dynamics},}\ }\href@noop {} {\bibfield
		{journal} {\bibinfo  {journal} {Elsevier Science}\ }\textbf {\bibinfo {volume} {6}}
		(\bibinfo {year} {2013})}\BibitemShut {NoStop}%
		\bibitem [{\citenamefont {Steinhauer}\ and\ \citenamefont {Ishida}(1997)}]{S1997}%
	\BibitemOpen
	\bibfield  {author} {\bibinfo {author} {\bibfnamefont {L. C.}\ \bibnamefont
			{Steinhauer}}\ and\ \bibinfo {author} {\bibfnamefont {A.}\
			\bibnamefont {Ishida}},\ }\bibfield  {title} {\enquote {\bibinfo {title}
			{Relaxation of a Two-Specie Magnetofluid},}\ }\href@noop {} {\bibfield
		{journal} {\bibinfo  {journal} {Physical Review Letters}\ }\textbf
		{\bibinfo {volume} {79}},\ \bibinfo {pages} {3423--3426} (\bibinfo {year}
		{1997})}\BibitemShut {NoStop}%
	\bibitem [{\citenamefont {Andr\'es}\ \emph {et~al.}(2014)\citenamefont
		{Andr\'es}, \citenamefont {Martin}, \citenamefont {Dmitruk}, 
		\ and\ \citenamefont {G\'omez}}]{Andres14}%
	\BibitemOpen
	\bibfield  {author} {\bibinfo {author} {\bibfnamefont {N.}~\bibnamefont
			{Andr\'es}}, \bibinfo {author} {\bibfnamefont {L.}~\bibnamefont {Martin}}, \bibinfo {author} {\bibfnamefont {P.}~\bibnamefont {Dmitruk}}, \ and\ \bibinfo {author}
		{\bibfnamefont {D.}\ \bibnamefont {G\'omez}},\ }\bibfield  {title}
	{\enquote {\bibinfo {title} {Effects of electron inertia in collisionless magnetic reconnection},}\ }\href {\doibase
		10.1063/1.4890021} {\bibfield  {journal} {\bibinfo  {journal}
			{Physics of Plasmas}\ }\textbf {\bibinfo {volume} {21}},\ \bibinfo
		{pages} {072904} (\bibinfo {year} {2014})}\BibitemShut {NoStop}%
		\bibitem [{\citenamefont {Comisso}\ \emph {et~al.}(2018)\citenamefont
		{Comisso}, \citenamefont {Huang}, \citenamefont {Lingam}, \citenamefont
		{Hirvijoki},\ and\ \citenamefont {Bhattacharjee}}]{Comisso18}%
	\BibitemOpen
	\bibfield  {author} {\bibinfo {author} {\bibfnamefont {L.}~\bibnamefont
			{Comisso}}, \bibinfo {author} {\bibfnamefont {Y.~M.}~\bibnamefont {Huang}},
		\bibinfo {author} {\bibfnamefont {M.}~\bibnamefont {Lingam}}, \bibinfo
		{author} {\bibfnamefont {E.}\ \bibnamefont {Hirvijoki}}, \ and\ \bibinfo {author}
		{\bibfnamefont {A.}\ \bibnamefont {Bhattacharjee}},\ }\bibfield  {title}
	{\enquote {\bibinfo {title} {Magnetohydrodynamic Turbulence in the Plasmoid-mediated Regime},}\ }\href {\doibase
		10.3847/1538-4357/aaac83} {\bibfield  {journal} {\bibinfo  {journal}
			{The Astrophysical Journal}\ }\textbf {\bibinfo {volume} {854}},\ \bibinfo
		{pages} {103} (\bibinfo {year} {2018})}\BibitemShut {NoStop}%
	\bibitem [{\citenamefont{Abdelhamid}\, \citenamefont{Kawazura}\ and\ \citenamefont{Yoshida}(2015)}]{Abdelhamid}%
	\BibitemOpen
	\bibfield  {author} {\bibinfo {author} {\bibfnamefont {H. M.}\ \bibnamefont
			{Abdelhamid}}\,   {\bibfnamefont {Y}\ \bibnamefont
			{Kawazura}}\ and\ \bibinfo {author} {\bibfnamefont {Z.}\
			\bibnamefont {Yoshida}},\ }\bibfield  {title} {\enquote {\bibinfo {title}
			{Hamiltonian formalism of extended magnetohydrodynamics},}\ }\href@noop {} {\bibfield
		{journal} {\bibinfo  {journal} {Journal of Physics A: Mathematical and Theoretical}\ }\textbf
		{\bibinfo {volume} {48}},\ \bibinfo {pages} {235502} (\bibinfo {year}
		{2015})}\BibitemShut {NoStop}%
		\bibitem [{\citenamefont {Schmidt}(2014)}]{Schmidt}%
	\BibitemOpen
	\bibfield  {author} {\bibinfo {author} {\bibfnamefont {W.}~\bibnamefont
			{Schmidt}},\ }\href@noop {} {\emph {\bibinfo {title} {Numerical Modelling of astrophysical turbulence}}}\ (\bibinfo  {publisher} {Springer Briefs in astronomy},\ \bibinfo {year} {2014})\BibitemShut {NoStop}%
				\bibitem [{\citenamefont {Miura}(1982)}]{Miura}%
	\BibitemOpen
	\bibfield  {author} {\bibinfo {author} {\bibfnamefont {A.}\ \bibnamefont
			{Miura}},\ }\bibfield  {title} {\enquote {\bibinfo {title} {Nonlinear Evolution of the Magnetohydrodynamic Kelvin-Helmholtz Instability},}\ }\href@noop {} {\bibfield
		{journal} {\bibinfo  {journal} {Physical review letters}\ }\textbf {\bibinfo
			{volume} {49}},\ \bibinfo {pages} {779--782} (\bibinfo {year}
		{1982})}\BibitemShut {NoStop}%
\bibitem [{\citenamefont {Comisso}\ \emph {et~al.}(2016)\citenamefont
		{Comisso}, \citenamefont {Lingam}, \citenamefont {Huang}, 
		\ and\ \citenamefont {Bhattacharjee}}]{Comisso16}%
	\BibitemOpen
	\bibfield  {author} {\bibinfo {author} {\bibfnamefont {L.}~\bibnamefont
			{Comisso}}, \bibinfo {author} {\bibfnamefont {M.}~\bibnamefont {Lingam}}, \bibinfo {author} {\bibfnamefont {Y.~M.}~\bibnamefont {Huang}}, \ and\ \bibinfo {author}
		{\bibfnamefont {A.}\ \bibnamefont {Bhattacharjee}},\ }\bibfield  {title}
	{\enquote {\bibinfo {title} {General theory of the plasmoid instability},}\ }\href {\doibase
		10.1063/1.4964481} {\bibfield  {journal} {\bibinfo  {journal}
			{Physics of Plasmas}\ }\textbf {\bibinfo {volume} {23}},\ \bibinfo
		{pages} {100702} (\bibinfo {year} {2016})}\BibitemShut {NoStop}%

	
	
\end{thebibliography}
\end{document}